\documentclass[tightenlines,twoside,secnumarabic,
               onecolumn,floatfix,nofootinbib,preprintnumbers,11pt]{revtex4}
\usepackage{graphicx}
\usepackage{dcolumn}
\usepackage{float}
\usepackage{amsmath}
\usepackage{amsfonts}
\usepackage{multirow}
\usepackage{color}
\usepackage{wasysym}
\usepackage{soul}
\usepackage{slashed}

\def\GeV{\ifmmode {\mathrm{\ Ge\kern -0.1em V}}\else
                   \textrm{Ge\kern -0.1em V}\fi}%
\def\TeV{\ifmmode {\mathrm{\ Te\kern -0.1em V}}\else
                   \textrm{Te\kern -0.1em V}\fi}%

\newcommand{\beq}{\begin{equation}}
\newcommand{\eeq}{\end{equation}}
\newcommand{\beqn}{\begin{eqnarray}}
\newcommand{\eeqn}{\end{eqnarray}}

\newcommand{\nn}{\nonumber}

\newcommand{\fb}{{\ensuremath\rm fb}}
\newcommand{\ora}{\overrightarrow}

\newcommand{\leftb}{\left(}
\newcommand{\rightb}{\right)}

\def\beq{\begin{equation}}
\def\eeq{\end{equation}}
\def\beqn{\begin{eqnarray}}
\def\eeqn{\end{eqnarray}}
\def\nn{\nonumber}
\def\spa#1.#2{\left\langle#1\,#2\right\rangle}
\def\spb#1.#2{\left[#1\,#2\right]}
\def\spaa#1.#2.#3{\langle\mskip-1mu{#1} 
                  | #2 | {#3}\mskip-1mu\rangle}
\def\spbb#1.#2.#3{[\mskip-1mu{#1}
                  | #2 | {#3}\mskip-1mu]}
\def\spab#1.#2.#3{\langle\mskip-1mu{#1} 
                  | #2 | {#3}\mskip-1mu]}
\def\spba#1.#2.#3{[\mskip-1mu{#1} 
                  | #2 | {#3}\mskip-1mu\rangle}
\def\spaba#1.#2.#3.#4{\langle\mskip-1mu{#1} 
                  | #2 | #3 | {#4}\mskip-1mu\rangle}

\def\bentarrow{\:\raisebox{1.3ex}{\rlap{$\vert$}}\!\rightarrow}
\def\bothdk#1#2#3#4#5{
\begin{array}{r c l}
#1 & \rightarrow & #2#3 \\
 & & \:\raisebox{1.3ex}{\rlap{$\vert$}}\raisebox{-0.5ex}{$\vert$}
\phantom{#2}\!\bentarrow #4 \\
 & & \bentarrow #5
\end{array}
}

\begin{document}

\preprint{FERMILAB-PUB-15-257-T}

\title{Next-to-leading order predictions for $WW$+jet production}

\author{John M. Campbell}
\affiliation{%
Fermilab, PO Box 500, Batavia, IL 60510, USA}

\author{David J. Miller}
\affiliation{%
School of Physics and Astronomy, University of Glasgow, Glasgow, G12 8QQ, UK}

\author{Tania Robens}
\affiliation{%
IKTP, Technische Universit\"at Dresden, Zellescher Weg 19, D-01069, Dresden, Germany \\}

\date{\today}

\begin{abstract}
In this work we report on a next-to-leading order calculation of $WW$ + jet production at hadron colliders,
with subsequent leptonic decays of the $W$-bosons included.  The calculation of the one-loop
contributions is performed using generalized unitarity methods in order to derive analytic expressions
for the relevant amplitudes.  These amplitudes have been implemented in the parton-level Monte Carlo 
generator MCFM, which we use to provide a complete next-to-leading order calculation. Predictions for
total cross-sections, as well as differential distributions for several key observables, are computed
both for the LHC operating at $14$~TeV as well as for a possible future $100$~TeV proton-proton collider.

\end{abstract}

\pacs{Valid PACS appear here}
\maketitle

\section{Introduction}

In this paper we report on a calculation of the next-to-leading order (NLO) QCD corrections to the
process of $W$ pair production in association with a jet using unitarity methods, taking into account
spin correlations in the leptonic decays of the $W$ bosons.
This process is important for a number of reasons.
Firstly, the rate for $WW$ production at the LHC is significant and thus the $WW$ process (and more generally all
the vector boson pair production processes) provide a useful laboratory with which to probe the Standard Model (SM).
Indeed, already in Run I of the LHC, the ATLAS and CMS
experiments have been able to investigate the properties of the $WW$ process in some detail~\cite{ATLAS:2012mec,
ATLAS-CONF-2014-033,Chatrchyan:2013oev,Aad:2014mda,Aad:2014jra,CMS-PAS-SMP-14-016, Chatrchyan:2013yaa,Chatrchyan:2011tz,Chatrchyan:2012bd}.
The presence of an additional jet in the detector acceptance only slightly reduces
the cross-section, by around a factor of $2$--$3$ for typical jet cuts.
In addition, this process also represents an important background to the production of
a Higgs boson with subsequent decay into $W$ pairs, either through the
gluon-fusion channel with an additional jet present, or through weak boson
fusion where only one of the forward jets is detected.
Analyses of $WW$ events also provide strong constraints on anomalous triple and quartic
gauge couplings~\cite{ATLAS:2012mec,CMS-PAS-SMP-14-016,Chatrchyan:2013yaa,Khachatryan:2014kca} and 
represent important backgrounds for searches for additional scalars of higher
mass~\cite{ATLAS-CONF-2013-067, Khachatryan:2015cwa}.

Aside from the immediate relevance to the LHC experimental program, the
one-loop amplitudes for the process at hand represent an important component
of the NNLO corrections to the $WW$ process.  Indeed, calculations of the corresponding two-loop
amplitudes~\cite{Chachamis:2008yb,Chachamis:2008xu,Gehrmann:2014fva,vonManteuffel:2015msa,Caola:2015ila,Gehrmann:2015ora} have already allowed
first determinations of the NNLO contribution~\cite{Gehrmann:2014fva}.
In such a calculation the one-loop $WW + 3$~parton amplitudes must be evaluated in the limit in which the gluon
is soft, or the quark and antiquark are collinear.  This is true for both the sector-decomposition
and antenna-subtraction methods that have mostly been employed in NNLO calculations thus far and
also for the recently-introduced SCET-based $N$-jettiness method~\cite{Boughezal:2015dva,Gaunt:2015pea}.
For this reason it is important that the form of the amplitudes be both numerically stable and
efficiently evaluated.

In this paper we evaluate the NLO corrections to the $WW$+jet process, using analytic unitarity
methods~\cite{Britto:2004nc,Britto:2005ha,Britto:2006sj,Forde:2007mi,Mastrolia:2009dr,Badger:2008cm}
in order to obtain the necessary one-loop amplitudes.
This technique allows the amplitude to be represented in a compact analytic form. Similar calculations have
provided one-loop amplitudes for a wealth of important processes such as the production of Higgs$+2$~jets~\cite{Dixon:2009uk,
Badger:2009hw, Badger:2009vh}, $t\bar{t}$~\cite{Badger:2011yu},
$Wb{\bar b}$~\cite{Badger:2010mg}, and three-\cite{Campbell:2014yka} and four-photon
production~\cite{Dennen:2014vta}. The implementation of analytic expressions for the
virtual contributions addresses both of the issues previously mentioned: it is relatively compact and 
improves numerical stability.  As a result such calculations have already been employed in NNLO calculations
of the Higgs+jet~\cite{Boughezal:2013uia,Chen:2014gva,Boughezal:2015dra,Boughezal:2015aha} and $W$+jet~\cite{Boughezal:2015dva} processes.

Previous calculations of the $WW$+jet process have
employed either a numerical OPP procedure~\cite{Campbell:2007ev}, 
variants of traditional integral reduction
methods~\cite{Dittmaier:2007th,Sanguinetti:2008xt,Dittmaier:2009un} or a combination
of the two strategies~\cite{Cascioli:2013gfa}.
Furthermore, results for the $WW$+jet process at NLO have already been made
available through the MCFM+ code~\cite{Melia:2012zg}, including also gluon-induced processes that
formally constitute a NNLO contribution.  Here we improve on the implementation of the NLO
corrections  by providing much faster code directly in the main release of MCFM~\cite{Campbell:1999ah},
in such a manner that it is also able to  take advantage of the recent multi-threaded improvement
to the code~\cite{Campbell:2015qma}.

\section{Calculation}

\begin{figure}
\begin{center}
\includegraphics[scale=0.4]{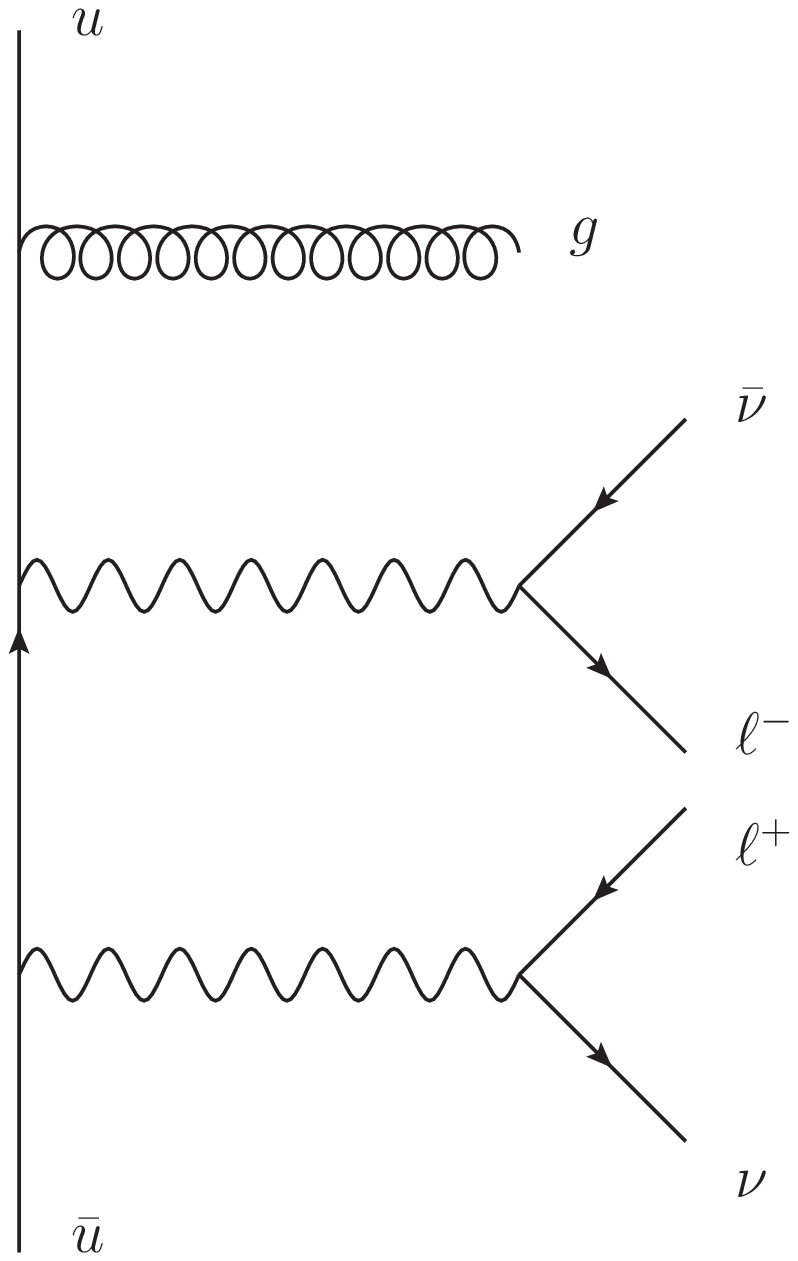} \hspace*{1.5cm}
\includegraphics[scale=0.4]{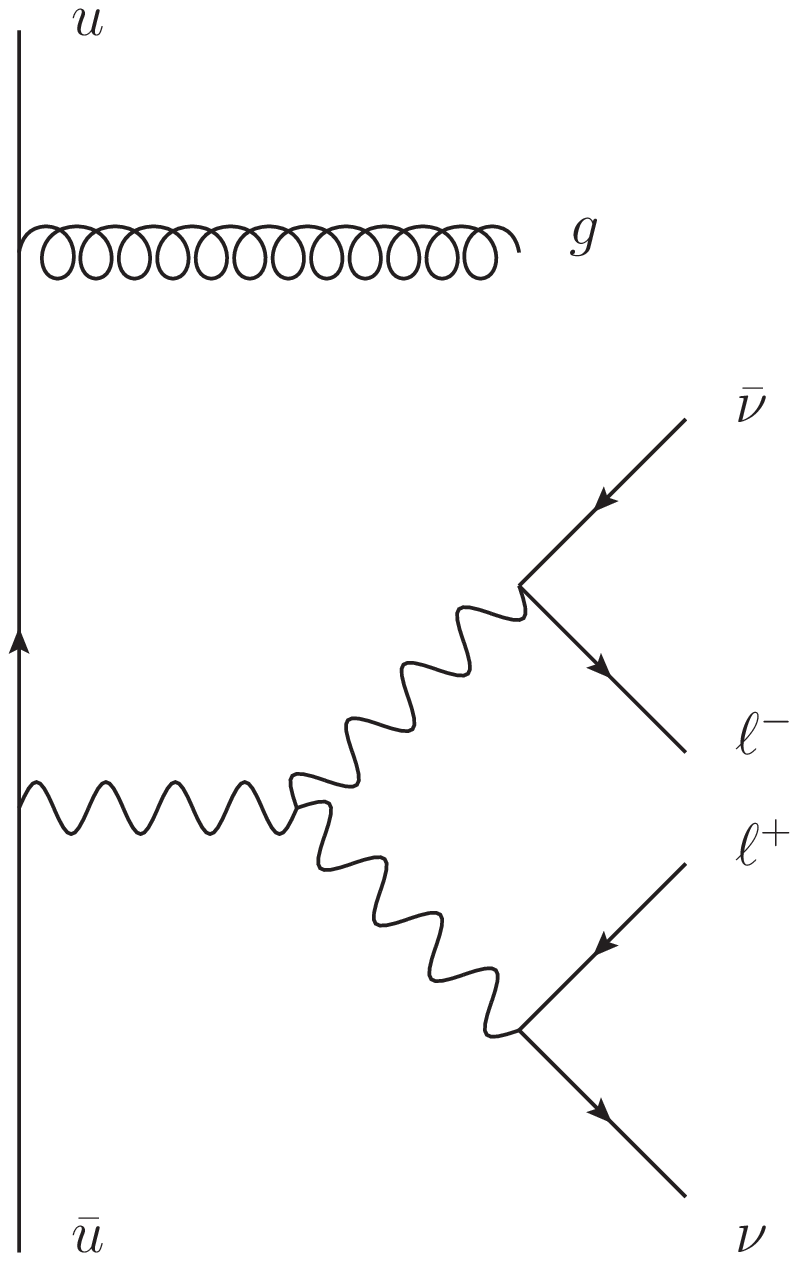} \\ ~~ \\
(a) \hspace*{5cm} (b)
\end{center}
\caption{Sample diagrams entering the calculation of the leading order amplitude
for the $WW+$jet process, corresponding to (a) $W$ emission from the quark line and (b) emission
from an intermediate $Z$ boson or photon.\label{fig:LOdiags}}
\end{figure}

In this paper we will consider the hadronic production of $W$ pairs in association with a single
jet.  The $W$ bosons decay leptonically, with all spin correlations included.  At tree level
this process corresponds to the partonic reaction,
\begin{equation}
\label{WWjetprocess}
\bothdk{q+\bar q}{W^+ +}{W^-+g}{\mu^-+\nu_\mu}{\nu_e+e^+}
\end{equation}
with all possible crossings of the partons between initial and final states.  Although we always
include the leptonic decays of the $W$ bosons, for brevity we will refer to this as the $WW$+jet process.
At tree level this reaction proceeds through Feynman diagrams where both $W$s are directly emitted
from the quark line, as well as diagrams where the $W$ pair stems from an intermediate
$\gamma$ or $Z$.  Representative diagrams for these are shown in
Fig.~\ref{fig:LOdiags}.  Expressions for the tree-level amplitudes have been presented
previously~\cite{Dixon:1998py,Campbell:2007ev} and these have been implemented in our calculation.

At next-to-leading order we must include the emission of an additional parton, either as a virtual
particle to form a loop amplitude, or as a real external particle.  For the latter contribution
we use the matrix elements previously computed in Ref.~\cite{Campbell:2007ev} and employ the
Catani-Seymour dipole subtraction method~\cite{Catani:1996vz} to isolate the soft and collinear
singularities in dimensional regularization.  The computation of the one-loop amplitudes in analytic
form is highly non-trivial and is the central result of this paper.
These contributions have been implemented in the code MCFM, so that complete
NLO predictions for both total cross-sections and differential distributions are readily
available.

\begin{figure}
\begin{center}
\includegraphics[scale=0.4]{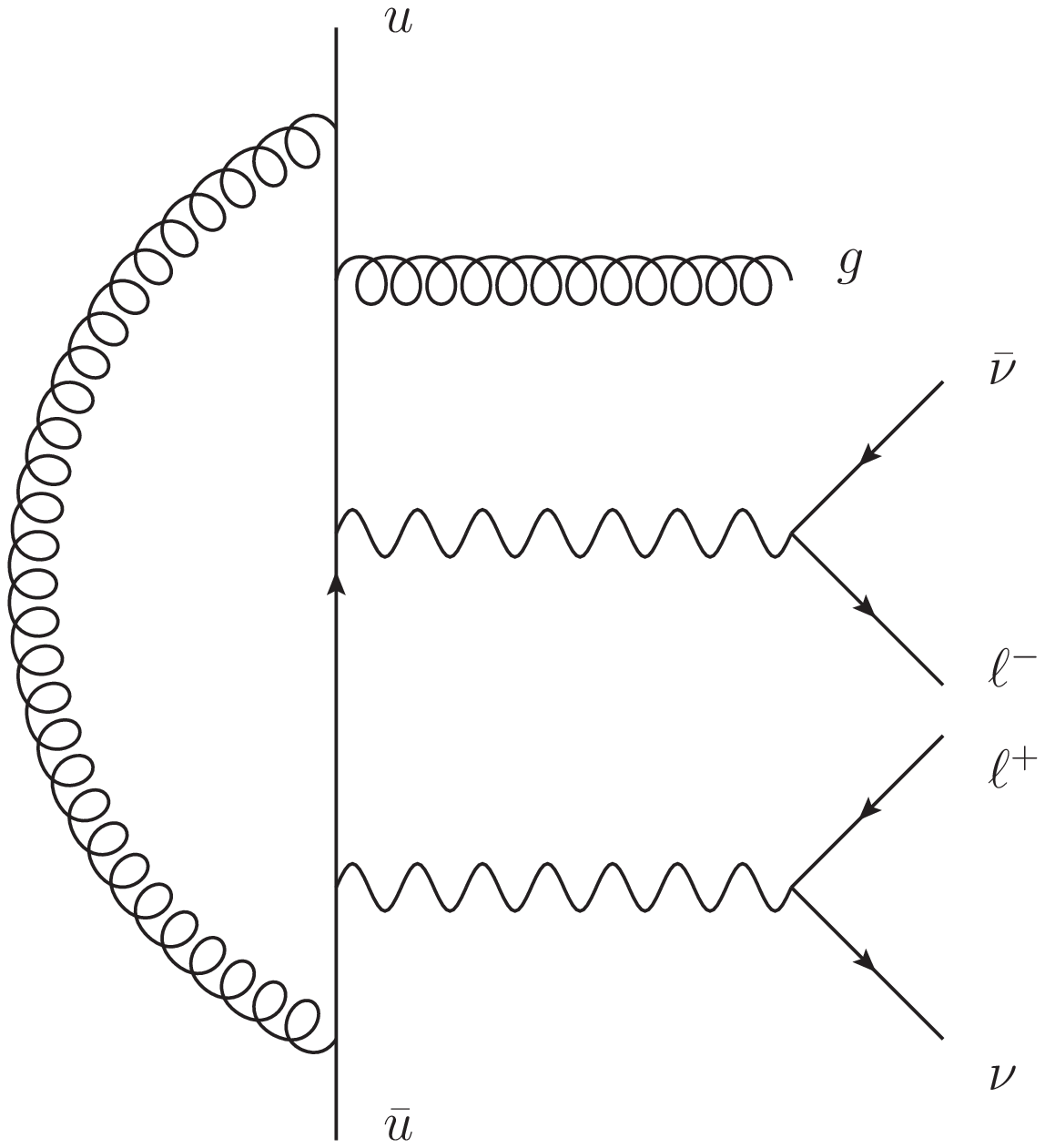} \hspace*{1.5cm}
\includegraphics[scale=0.4]{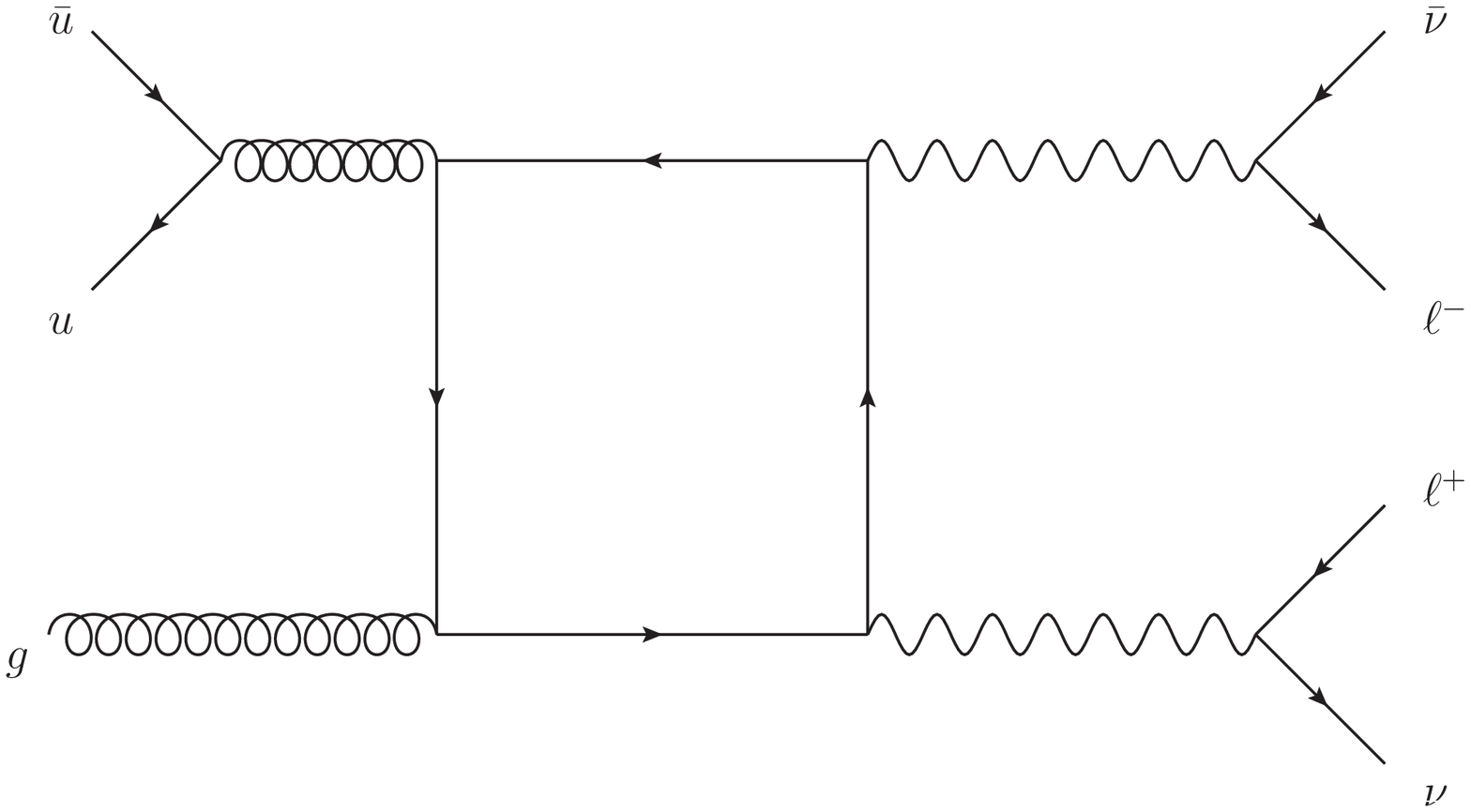}
\end{center}
\caption{Sample diagrams entering the calculation of the one-loop amplitude
for the $WW+$jet process.  The one-loop diagrams can be categorized according
to whether a gluon dresses a leading-order amplitude (left), or whether the
diagram includes a closed fermion loop (right).\label{fig:loopdiags}}
\end{figure}
The one-loop corrections to diagrams such as the one shown in Fig.~\ref{fig:LOdiags}(b) are relatively
straightforward to compute.  Rather than computing them explicitly, the contribution is obtained by
recycling existing results for $q\,\bar{q} \to \,Z(\to \ell^- \ell^+) + g$~\cite{Bern:1997sc},
with the leptonic $Z$ decay current replaced by the one for $Z \to W^+ W^- \to \ell^+ \nu \ell^- \bar\nu$. 
The two remaining classes of diagrams to consider are depicted in Fig.~\ref{fig:loopdiags} and correspond
to either an internal gluon propagator dressing of Fig.~\ref{fig:LOdiags}(a) or diagrams that
contain a closed fermion loop.  The amplitude representing the closed fermion loops
is finite, which can easily be seen from the lack of a tree-level $ggW^+W^-$ coupling in the SM.
Both of these sub-amplitudes are computed using generalized unitarity
methods as follows.  Each amplitude is decomposed in terms of the usual one-loop basis of
box, triangle and bubble integrals, i.e.
\begin{eqnarray}
\mathcal{A}(\{p_i\}) = \sum_{j} d_{j} I_4^j +  \sum_{j} c_{j} I_3^j +  \sum_{j} b_{j} I_2^j + R \;.
\label{eq:ampdecomp}
\end{eqnarray}
In this equation $I_n^j$ represents a scalar loop integral with $n$ propagators, commonly referred
to as box ($n=4$), triangle ($n=3$) and bubble ($n=2$) integrals.  The integral coefficients
$d_j$, $c_j$ and $b_j$ can be obtained by the application of unitarity cuts in
four dimensions~\cite{Britto:2004nc,Britto:2005ha,Britto:2006sj,Forde:2007mi,Mastrolia:2009dr}.
The rational remainder term $R$ can be determined using similar cutting rules, after the inclusion
of a fictitious mass for the particles propagating in the loop~\cite{Badger:2008cm}.  Since the
tree-level on-shell amplitudes that appear in the cutting procedure are quite complex, this procedure
has been performed using the help of the S@M Mathematica package~\cite{Maitre:2007jq}.
The evaluation of the scalar integrals appearing in Eq.~(\ref{eq:ampdecomp}) has been performed with
the aid of the QCDLoop Fortran library~\cite{Ellis:2007qk}.

\subsection{Details}

In this section we will present a number of the integral coefficients that were computed using
the unitarity techniques described previously.  For the sake of brevity we will show coefficients
that do not involve lengthy algebraic manipulation and are therefore particularly compact.  The remaining
coefficients can be inspected in the distributed MCFM code.

\begin{figure}
\begin{center}
\includegraphics[scale=0.4]{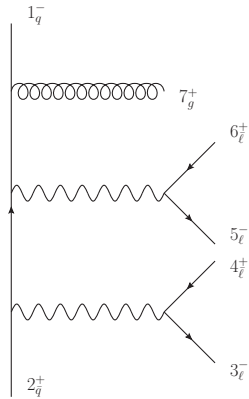} 
\end{center}
\caption{The notation used in the calculation of the $WW+$jet process, corresponding to Eq.~(\ref{eq:treeproc}).
\label{fig:notation}}
\end{figure}

In order to establish the notation for these coefficients
we first discuss the corresponding leading-order results.  We consider all particles
outgoing and consider the process,
\beq\label{eq:treeproc}
0 \to q^-(p_1) + {\bar q}^+(p_2) + \ell^-(p_3) + \bar \ell^+(p_4) + \ell^-(p_5) + \bar\ell^+(p_6) + g^+(p_7) \;,
\eeq
also shown in Figure~\ref{fig:notation},
where momentum assignments are shown in parentheses and superscripts denote the particle
helicities and polarizations.
This is the basic configuration for which we compute all amplitudes; other helicity configurations
are obtained by means of charge conjugation and parity relations.  Tree-level amplitudes for this process
have been presented in detail in Refs.~\cite{Dixon:1998py, Campbell:2007ev}, whose notation we follow
closely.  The full amplitude can be written in terms of two primitive amplitudes,
$A^{(a)}$ and $A^{(b)}$ corresponding to diagrams of the two types depicted in Figure~\ref{fig:LOdiags}.
The full amplitude is obtained from the primitive ones by dressing with appropriate color and
coupling factors~\cite{Dixon:1998py}.
The explicit forms, for the assignment of momenta shown in Eq.~(\ref{eq:treeproc}), are:
\begin{eqnarray*}
\lefteqn{A^{(a)}(1_{q}^{-},2_{\bar{q}}^{+},\,3_{l}^{-},\,4_{\bar{l}}^{+},\,5_{l}^{-}\,6_{\bar{l}}^{+}\,,7_{g}^{+})}\\
&=&\frac{1}{s_{ 34}-m_{W}^{2}}\,\frac{1}{s_{ 56}-m_{W}^{2}}\,\frac{1}{s_{156}}\,
\frac{\langle  1 5\rangle}{\langle  1 7\rangle}
\Bigg\{\frac{[ 6|1+ {5}|3\rangle\,[ 4| { 2}+ {7}| 1\rangle}{\langle  2 7\rangle}
\,+\,\frac{[ 6 5]\langle 5  1 \rangle\,[7| { 2}+ { 4}|3\rangle\,[ 4  2]}{s_{ 2 3 4}}\Bigg\},\\
&&\\
\lefteqn{A^{(b)}(1_{q}^{-},2_{\bar{q}}^{+},\,3_{l}^{-},\,4_{\bar{l}}^{+},\,5_{l}^{-}\,6_{\bar{l}}^{+}\,,7_{g}^{+})
\,=\,\frac{1}{s_{ 34}-m_{W}^{2}}\,\frac{1}{s_{ 56}-m_{W}^{2}}\,\frac{1}{s_{127}}  \frac{1}{\langle 17\rangle \langle 27\rangle}\times}\\
&& \Bigg\{\langle 15\rangle \langle 1 |2+7|6] \langle 3 |5+6|4]-\langle 13\rangle \langle 1 |2+7|4] \langle 5 |3+4|6 ] 
-\langle 35\rangle [46] \langle 1 |(3+4)(2+7)|1 \rangle 
\Bigg\}.
\end{eqnarray*}

The coefficients discussed below all appear in the amplitude representing the diagrams
of Figure~\ref{fig:loopdiags} (left), i.e. they do not contain a closed loop of fermions.
The decomposition of the amplitude into the form shown in Eq.~(\ref{eq:ampdecomp}) contains five
basis integrals corresponding to  ``three-mass'' boxes, i.e. four-point integrals with 
three non-lightlike external legs.  This naturally leads to a plethora of three-mass triangles,
whose coefficients do not lend themselves easily to a compact representation in terms of
straightforward spinor products.  As a result, many of the coefficients are considerably
more complex than those presented here.  To simplify the calculation slightly, we use the
known singular structure of the amplitude to determine one of the bubble coefficients
from the remainder.  

As a representative box integral coefficient we choose the one
corresponding to the basis integral
$I_4 \leftb s_{56}, s_{34},0, s_{17};s_{127},s_{234}\rightb$.
We here show the leading color integral coefficient, which receives a pre-factor
of $N_c$.  It can be written as, 
\begin{eqnarray}
\lefteqn{d\leftb s_{56}, s_{34},0, s_{17};s_{127},s_{234}\rightb = 
\frac{1}{s_{34}-m_W^2}\,\frac{1}{s_{56}-m_W^2}\frac{\langle  1 2\rangle^2 \,\langle 2|P|2 ]}{2 \, \langle  2 7 \rangle\,\langle 1 7 \rangle} \times} \nn\\
&& \leftb  [{4} {2}]-\frac{\langle {2}|P|{4}] }{D}\rightb \leftb  \langle
   {3}|{2+4}|{6}] -\frac{ \langle {2} {3}\rangle\langle {2}|P|{6}] }{D} \rightb\,\leftb \frac{[{7} {1}] \langle {1} {5}\rangle  }{\langle {2}|P|{7}]}+\frac{ \langle {2} {5}\rangle}{D} \rightb 
\end{eqnarray}
where the compound momentum $P$ and denominator factor $D$ are defined by,
\begin{eqnarray}
P&=&s_{17}\,p_{34}+s_{234}p_{17},\, \qquad
D\,=\,\langle  2|( 3+{ 4})\,( { 1}+ {7})| 2\rangle.
\end{eqnarray}
The factors of $D$ can be put into a more familiar form by relating them to the product
$D D^\star$, where the complex conjugate of $D$ is simply given by $D^\star=[ 2|( 3+{ 4})\,( { 1}+ {7})| 2 ]$.
The product can be written as a trace of gamma matrices that evaluates to,
\begin{equation}
D D^\star = 
  4 s_{34} (p_2 \cdot p_{17} )^2 
+ 4 s_{17} (p_2 \cdot p_{34} )^2 
- 8 (p_2 \cdot p_{17}) (p_2 \cdot p_{34}) (p_{17} \cdot p_{34}) \;.
\label{eq:3boxgram}
\end{equation}
This is just the Gram determinant for this basis integral;  its presence,
when raised to a sufficiently high power, can lead to numerical instability
in phase space regions where it is very small. To avoid any such issues we
veto phase regions where cancellations between the terms in
Eq.~(\ref{eq:3boxgram}) (and equivalent expressions for the other
box integrals) occur at the level of $10^{-6}$ or more.
In our studies this occurs only very rarely, in about one in a million
events, so that the effects of such a veto are tiny
compared to the anticipated level of precision.

For the calculation of the integral coefficients in four dimensions, the only
triangle coefficients that must be computed correspond to integrals with three
massive external legs.  Triangle integrals with one or more lightlike legs only
contribute to the overall pole structure, which is known a priori.  The three mass
triangle coefficients are most easily expressed in terms of an extended set of
momenta that naturally appear in the unitarity approach~\cite{Forde:2007mi}.
For example, the coefficient of the (leading-colour)
basis integral $I_3(s_{34},s_{27},s_{156})$ is,
\begin{eqnarray}
c(s_{34},s_{27},s_{156})&=&
\frac{1}{2}\frac{1}{s_{34}-m_W^2}\,\frac{1}{s_{56}-m_W^2}\sum_{\gamma=\gamma_{1,2}}\frac{s_{ {2} {7}} [ {4} K_2^\flat] [ {7}  {2}] 
[ 6  {5}] \langle K^\flat_1  {2}\rangle  \langle 
K^\flat_1  {3}\rangle  \langle  {1}  {5}\rangle^2}{\left(\gamma-s_{ {2} {7}}\right) [ {7} K_2^\flat] \
\langle K^\flat_1  {1}\rangle  \langle K^\flat_1  {7}\rangle  \
\langle  {2}  {7}\rangle }
\end{eqnarray}
where the additional momenta $K^\flat_1$ and $K^\flat_2$ are defined by,
\begin{equation}
K^\flat_1\,=\, \frac{\gamma\left[ \gamma\, p_{27}+s_{ 2  7}\,p_{34} \right]}{\gamma^2-s_{ 2  7}\,s_{ 3  4}},\; \qquad
K^\flat_2\,=\, -\frac{\gamma\left[ \gamma\,p_{34}+s_{ 3  4}\,p_{27} \right]}{\gamma^2-s_{ 2  7}\,s_{ 3  4}}.\\
\end{equation}
The values of $\gamma$ appearing in these equations are determined by the
condition that $K^\flat_1$ and $K^\flat_2$ are lightlike, 
\begin{equation}
\gamma_{1,2} = p_{27}\,\cdot\,p_{34}\,\pm\,
 \sqrt{\leftb p_{27}\,\cdot\,p_{34}\rightb^2-s_{ 2  7}\,s_{ 3  4}} \;.
\end{equation}

The expressions for the bubble coefficients are, in general, rather complicated due to the
complexity of the tree amplitudes that appear either side of the cut.  However, the coefficient
of the bubble integral $I_2\,\leftb s_{156}\rightb$, that appears at leading colour, is rather
simple.  It is given by,
\begin{eqnarray}
&& b\leftb s_{156}\rightb \,=\,
\frac{1}{s_{34}-m_W^2}\,\frac{1}{s_{56}-m_W^2} \frac{\langle  5 6\rangle\,[ 4 3]}{\langle  2 7\rangle \, \langle 7|P| 1]}\,\times
 \Bigg\{ 
 -\frac{\langle  1 5\rangle [5  6] \langle 3|P|1]^{2}}{s_{156} \langle  1|P| 1]} 
\left[\frac{\langle  1 5\rangle [5  6]}{2 \langle  1|P| 1]}
+ \frac{\langle 7|P| 6]}{\langle7|P|1]} \right]  \nn \\ && \qquad
+ \frac{\langle 7 3\rangle^2 \langle 7|P| 6]\,[7  6]}{\langle 7|P|7]}
 \left[ \frac{1}{\langle 7|P|7]} \leftb \frac{\langle3|P|7]}{\langle 3 7\rangle}
 +\frac{[7  6] s_{156}}{ 2\,\langle 7|P| 6]} \rightb + \frac{\langle3|P|1]}{\langle 3 7\rangle\,\langle7|P|1]} \right] 
 \Bigg\}, 
\end{eqnarray}
where the momentum across the cut is $P=p_1+p_5+p_6$.

\subsection{Validation}

In order to verify the correctness of our calculation we have performed a variety
of cross-checks on both the one-loop amplitudes and the complete NLO calculation.
These consist of both independent calculations by other methods as well as
comparisons with results previously reported in the literature.

For the one-loop amplitude it is useful to perform cross-checks of the calculation
at single points in phase space.  All amplitudes have been cross-checked with an
independent numerical implementation of $D$-dimensional unitarity based on ref.~\cite{Ellis:2008ir},
which allows a verification of individual basis integral coefficients.  
We find agreement with both the results presented
in Table~3 (Appendix B) of Ref.~\cite{Campbell:2007ev} and those of Table 8 (Section 12.4)
of Ref.~\cite{Bern:2008ef} for the case of non-decaying $W$-bosons. 

To make a direct comparison at the level of the complete NLO cross-section, we have 
performed a calculation in the set-up of Ref.~\cite{Bern:2008ef} using 
aMC@NLO/Madgraph 5~\cite{Alwall:2011uj, Alwall:2014hca}.  We find complete agreement
between our result and Madgraph5, with integration errors for both being at the per mille level,
as shown in Table~\ref{MG5comparison}. The calculations of Ref.~\cite{Campbell:2007ev,Bern:2008ef} also include
contributions from top and bottom quarks for diagrams with internal quark loops,
as well as Higgs-induced diagrams for Ref.~\cite{Bern:2008ef}. In order to assess the effect of 
the loop of third generation quarks, and the Higgs boson,  we have artificially inflated
the top quark and Higgs boson masses in order to marginalize their effects on the NLO
cross-section. Variations of the total cross-section are within the
respective integration error, cf. Table~\ref{MG5comparison}.
Therefore, although these contributions are not included in our calculation, they
do not have a significant effect at this level. We find a small difference,
of about $0.75\%$, with the calculation of Ref.~\cite{Bern:2008ef}, as reported in Table~\ref{MG5comparison}.
We find larger differences with the published results of Ref.~\cite{Campbell:2007ev}, at the
level of a few percent.  Given the excellent agreement with the other available results,
we ascribe this to under-estimated Monte Carlo uncertainty in the earlier calculation.
\begin{center}
\begin{table}
\begin{tabular}{l|l|l}
calculation& parameters &$\sigma^\text{NLO}$~[pb] \\ \hline
MCFM       & default & 14.571\,(18)\\
MG5        & default &14.547\,(19)\\
MG5        & $m_h\,\times\,10, m_t\,\times\,10$&14.615\,(21)\\
MG5        & $m_h\,\times\,100, m_t\,\times\,100$&14.563\,(19) \\
DKU~\protect\cite{Dittmaier:2007th} & default & 14.678\,(10)
\end{tabular}
\caption{Total NLO cross-sections for the process $p\,p\,\rightarrow\,W^+\,W^-\,j$ at
the 14 \TeV~LHC, with parameter specifications as in Ref.~\cite{Bern:2008ef} (``default''), except
where noted otherwise. \label{MG5comparison}}
\end{table}
\end{center}

\section{Phenomenology}

\begin{table}
\begin{center}
\begin{tabular}{|c|c|c|c|}
\hline
$m_W$               & 80.385 GeV           & $\Gamma_W$ & 2.085 GeV \\
$m_Z$               & 91.1876 GeV          & $\Gamma_Z$ & 2.4952 GeV \\
$e^2$               & 0.095032             & $g_W^2$    & 0.42635 \\ 
$\sin^2\theta_W$    & $0.22290$            & $G_F$      & $0.116638\times10^{-4}$ \\
\hline
\end{tabular}
\caption{The values of the mass, width and electroweak parameters used to produce
the results in this paper.
\label{parameters}}
\end{center}
\end{table}
The results presented in this section have been obtained using the parameters shown in
Table~\ref{parameters}.  Note that we do not include the effects of any third-generation quarks,
either as external particles or in internal loops.  Since the effects of a non-diagonal CKM
matrix are very small, we also do not include them here.
In calculations of LO quantities we employ the CTEQ6L1 PDF set~\cite{Pumplin:2002vw},
while at NLO we use CT10~\cite{Lai:2010vv}.  The renormalization and factorization scales are
usually chosen to be the same, $\mu_R = \mu_F = \mu$, with our default scale choice
$\mu = \mu_0$ given by,
\begin{equation}
\mu_0 \equiv \frac{H_T}{2} = \frac{1}{2} \sum_i p_{\perp}^i \;.
\end{equation}
The sum over the index $i$ runs over all final state leptons and partons.
This choice of scale captures some of the dynamics of the process in a way
that is missed in, for instance, a fixed scale choice $\mu = m_W$ or other
common event-by-event scales~\cite{Berger:2009ep}. 
Jets are defined using the anti-$k_T$ algorithm with separation parameter
$R=0.5$ and must satisfy,
\begin{equation}
p_{\perp}^\text{jet} > 25~\mbox{GeV} \;, \qquad
|\eta^\text{jet}| <4.5 \;.
\label{eq:jetcuts}
\end{equation}

Since many phenomenological studies of this process have already been performed, both
at NLO~\cite{Campbell:2007ev,Dittmaier:2007th,Sanguinetti:2008xt,Dittmaier:2009un}
and including the effects of a parton shower at NLO~\cite{Cascioli:2013gfa}, in this
paper we restrict ourselves to a small number of pertinent applications.  To this end we
consider the immediate prospects in Run 2 of the LHC by presenting cross-sections at
$14$~TeV under a range of possible experimental cuts.  For a longer-term
view, we also consider the situation at a possible Future Circular Collider with proton-proton
collisions at $100$~TeV.

The total cross-sections for $WW$+jet production at these colliders
are collated in Table~\ref{xsecs}.  The effect of the decays of the $W$
bosons are not included and the jet is defined using the cuts given
in Eq.~(\ref{eq:jetcuts}).  The theoretical uncertainty is computed by using
a series of scale variations about the central choice $\mu_0$.  In order to properly
explore this uncertainty, we decouple $\mu_R$ and $\mu_F$ and consider their
variation separately.  The uncertainty corresponds to the most extreme predictions
for the four choices,
\begin{equation}
\left\{ \mu_R, \mu_F \right\} =
 \left\{ 2 \mu_0, 2\mu_0 \right\} ,
 \left\{ \mu_0/2, \mu_0/2 \right\} ,
 \left\{ 2 \mu_0, \mu_0/2 \right\} ,
 \left\{ \mu_0/2, 2\mu_0 \right\} \;.
\label{eq:scaleuncert} 
\end{equation}
At the LHC the uncertainty estimate corresponds to the first two variations
in Eq.~(\ref{eq:scaleuncert}), i.e. when the scales are varied together.
At $100$~TeV the last two scale variations are most important, due to
an accidental cancellation between the dependence on factorization and renormalization
scales when they are varied together in the same direction.
 At the LHC
this estimate of the uncertainty decreases from approximately $10\%$ at LO
to about $4\%$ at NLO.  At $100$~TeV the estimates of the uncertainty, both at LO
and NLO, are approximately a factor of two larger.
For reference, the corresponding NLO cross-sections for inclusive $WW$
production with the same input parameters are approximately $120$pb at
$14$~TeV and $1300$pb at $100$~TeV.
\renewcommand{\baselinestretch}{1.5}
\begin{table}
\begin{center}
\begin{tabular}{|r|c|c|}
\hline
$\sqrt s$~~~~ & $\sigma_{LO}$~[pb] & $\sigma_{NLO}$~[pb] \\
\hline
$13$~TeV      & $34.9_{-11.0\%}^{+11.4\%}$  & $42.9_{-3.7\%}^{+3.7\%}$ \\
$14$~TeV      & $39.5_{-11.0\%}^{+11.7\%} $  & $48.6_{-4.0\%}^{+3.8\%}$ \\
$100$~TeV     & $648_{-23.8\%}^{+22.3\%}   $  & $740_{-9.3\%}^{+4.5\%} $ \\
\hline
\end{tabular}
\renewcommand{\baselinestretch}{1.0}
\caption{Cross-sections for the process $p p \to WW$+jet at proton-proton colliders
of various energies, together with estimates of the theoretical uncertainty
from scale variation according to Eq.~(\ref{eq:scaleuncert}).
Monte Carlo uncertainties are at most a
single unit in the last digit shown shown in the table. 
\label{xsecs}}
\end{center}
\end{table}
\renewcommand{\baselinestretch}{1.0}

\begin{figure}
\begin{center}
\includegraphics[scale=0.45,angle=-90]{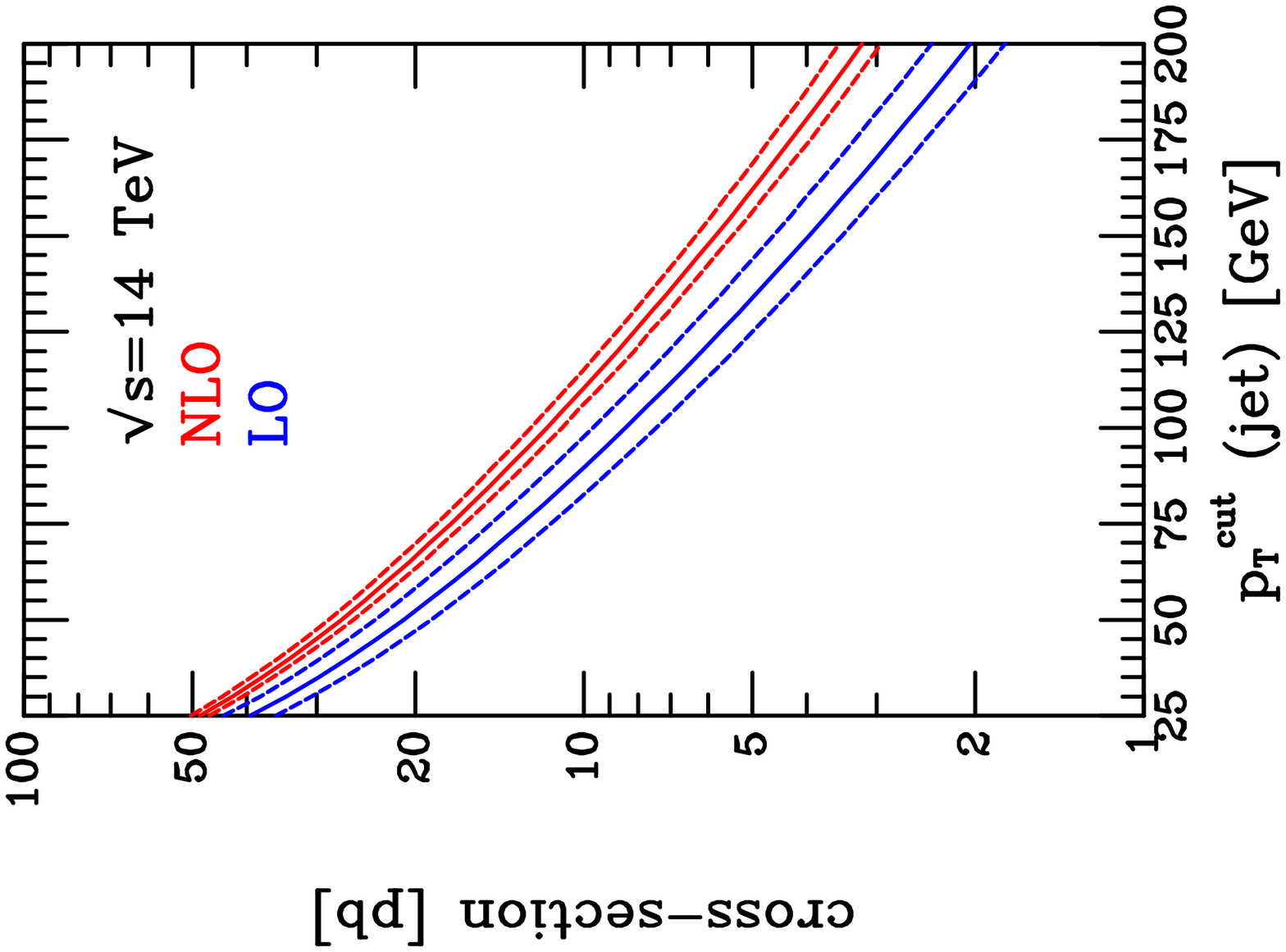} \hspace*{0.5cm}
\includegraphics[scale=0.45,angle=-90]{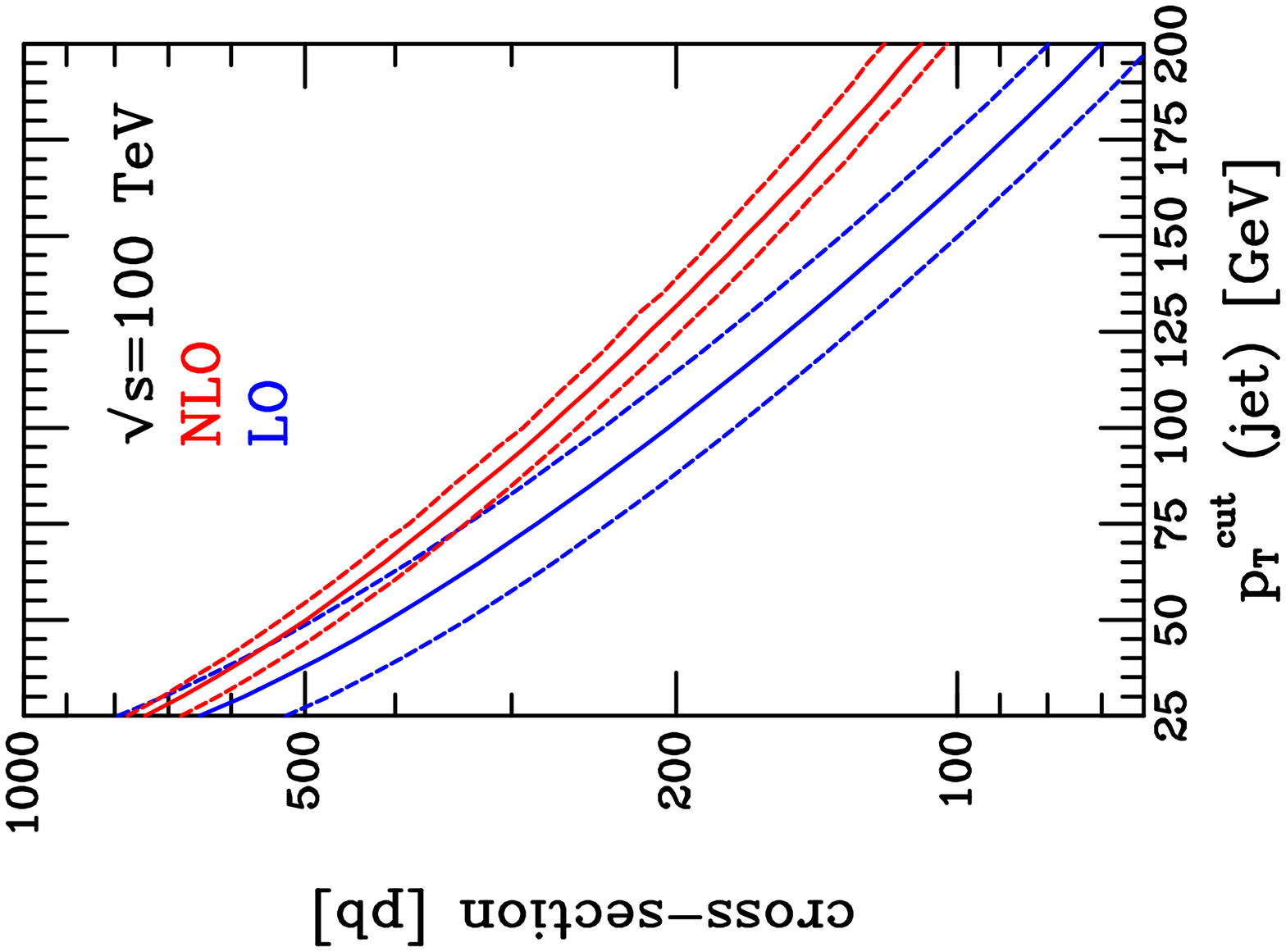}
\end{center}
\caption{Cross-sections at $\sqrt s = 14$~TeV (left) and $100$~TeV (right),
as a function of the transverse momentum
cut on the jet.  The prediction at each order is shown as a solid line,
with the dotted lines indicating the scale uncertainty corresponding to a factor of two variation
about the central scale.
\label{fig:ptjet}
}
\end{figure}

To assess the effect of jet cuts at higher transverse momenta, we also present the
cross-sections at LO and NLO as a function of the minimum jet $p_{\perp}$ in Figure~\ref{fig:ptjet}.
The size of the higher-order correction increases with the minimum jet $p_{\perp}$, although the relative
uncertainty is approximately the same. We note that the relative importance of the $WW+$jet final state,
compared to inclusive $WW$ production, is greater at the $100$~TeV collider.

We now turn away from more inclusive quantities and instead focus on particular
sets of cuts targetted at specific analyses.
We first consider the case of $14$~TeV LHC running, with a set of cuts
inspired by the ATLAS determination of the spin and parity of the Higgs boson
presented in Ref.~\cite{Aad:2015rwa}.  The $WW$ process constitutes the
largest irreducible background in the $H \to WW^\star$ decay channel and
a cocktail of cuts must be applied in order to access information about
the Higgs boson.  Our analysis is limited to the consideration of the
dominant $WW$+jet background and somewhat simplified
with respect to the experimental one.  The cuts that we apply
are summarized in Table~\ref{14TeVcuts}.  These include constraints on
the transverse mass of $(X,E^\text{miss}_T)$ systems, $m_T^X$,
where $X \in (\ell\ell, \ell_1, \ell_2)$, with
$p_{\ell\ell}\,=\,p_{\ell_1}+p_{\ell_2}$.  This quantity is defined
by\footnote{See, for instance, Eq.~(3) of Ref.~\cite{Khachatryan:2015cwa}.},
\begin{equation}
m_T^X\,=\,\sqrt{2\,p_{\perp}^X E^\text{miss}_T\,
 \left( 1-\cos\Delta\Phi(\ora{p}^X_T,\ora{E}_T^\text{miss})  \right)}.
\end{equation}
\begin{center}
\begin{table}
\begin{tabular}{l|l}
variable& cut \\ \hline
$p_{\perp,j}$&$>$ 25 \GeV\\
$|\eta_j|$&$<4.5$ \\
\hline
$|\eta_\ell|$&$<$ 2.5 \\
$p_{\perp,\ell_1}$&$>$ 22 \GeV \\
$p_{\perp,\ell_2}$&$>$ 15\,\GeV \\
$m_{\ell\ell}$&$\in[10, 80]\,\GeV$ \\
$p_\perp^\text{miss}$&$>$ 20\,\GeV \\
$\Delta\Phi_{\ell\ell}$&$<$ 2.8 \\
$m_T^{\ell\ell}$&$<$ 150 \GeV \\
$\text{max}[m_T^{\ell_{1}},m_T^{\ell_2}]$&$>$ 50 \GeV 
\end{tabular}
\caption{\label{14TeVcuts} 
Cuts applied in the $14$~TeV analysis, corresponding to
the ``full'' set of cuts.  The jet cuts, corresponding to the
first two lines in the table, are the only ones applied for
the ``basic'' cross-section.
}
\end{table}
\end{center}

\begin{figure}
\begin{minipage}{0.45\textwidth}
\includegraphics[height=\textwidth,angle=-90]{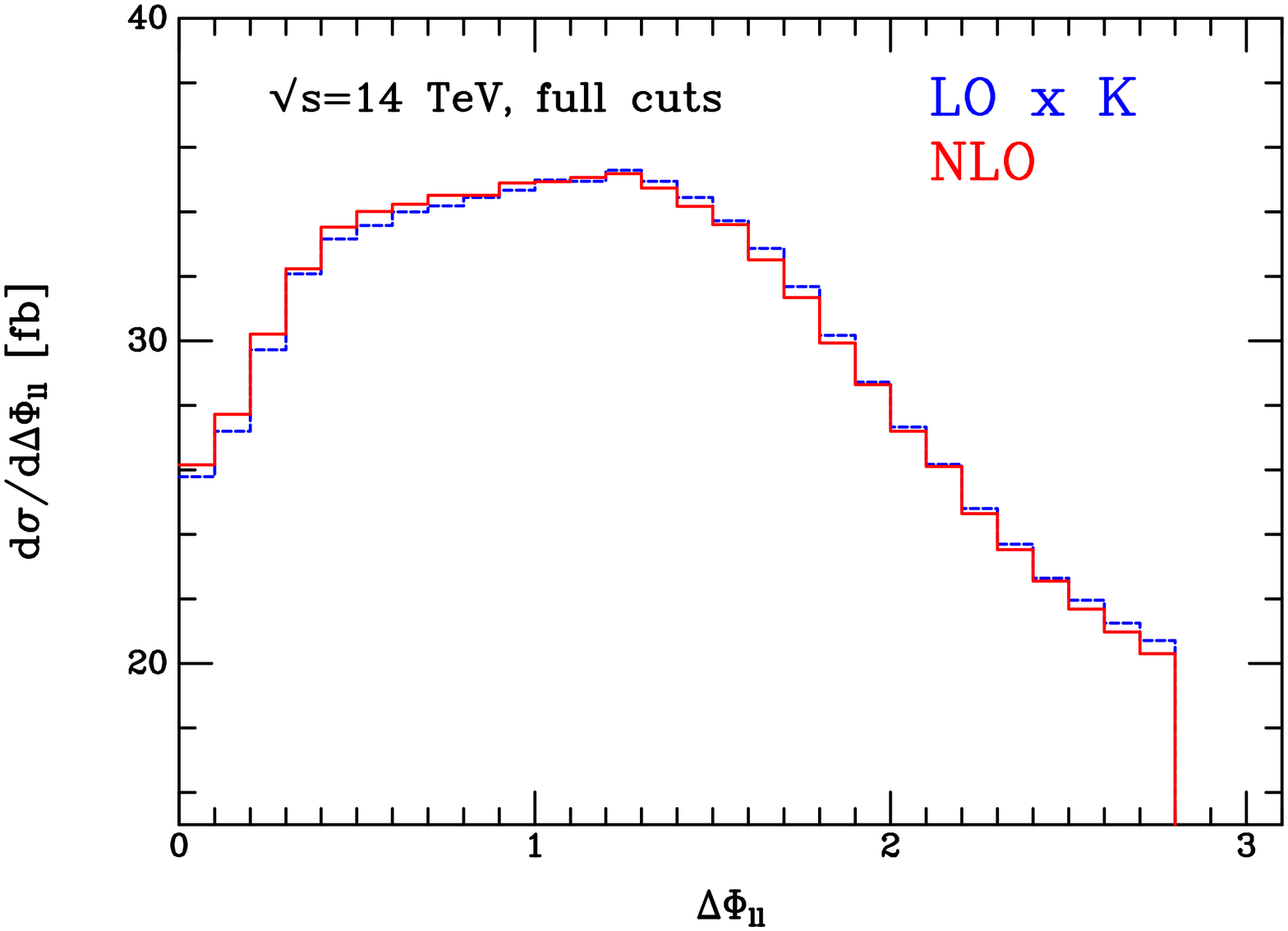}
\end{minipage}
\begin{minipage}{0.45\textwidth}
\includegraphics[height=\textwidth,angle=-90]{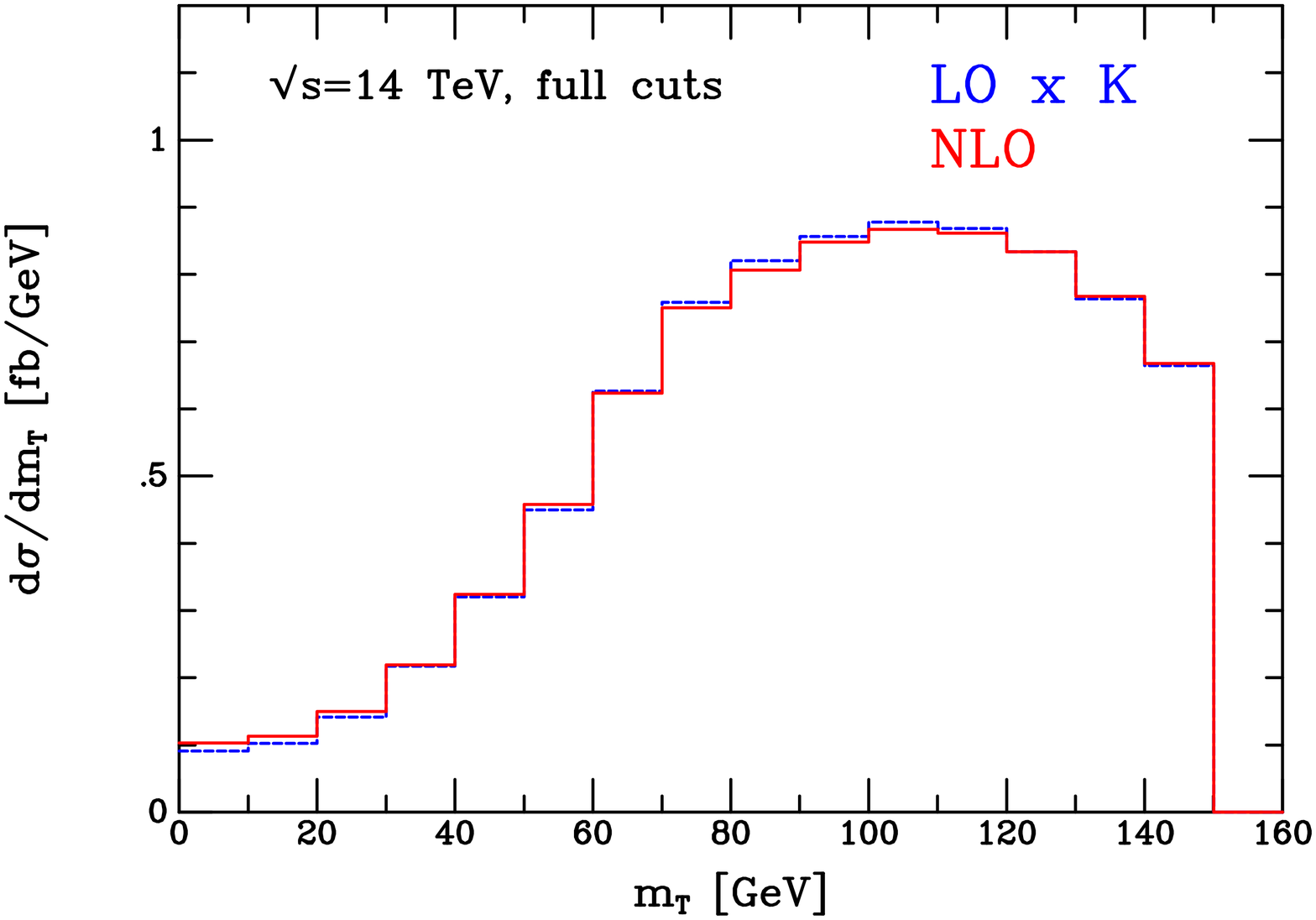}
\end{minipage}
\begin{minipage}{0.45\textwidth}
\includegraphics[height=\textwidth,angle=-90]{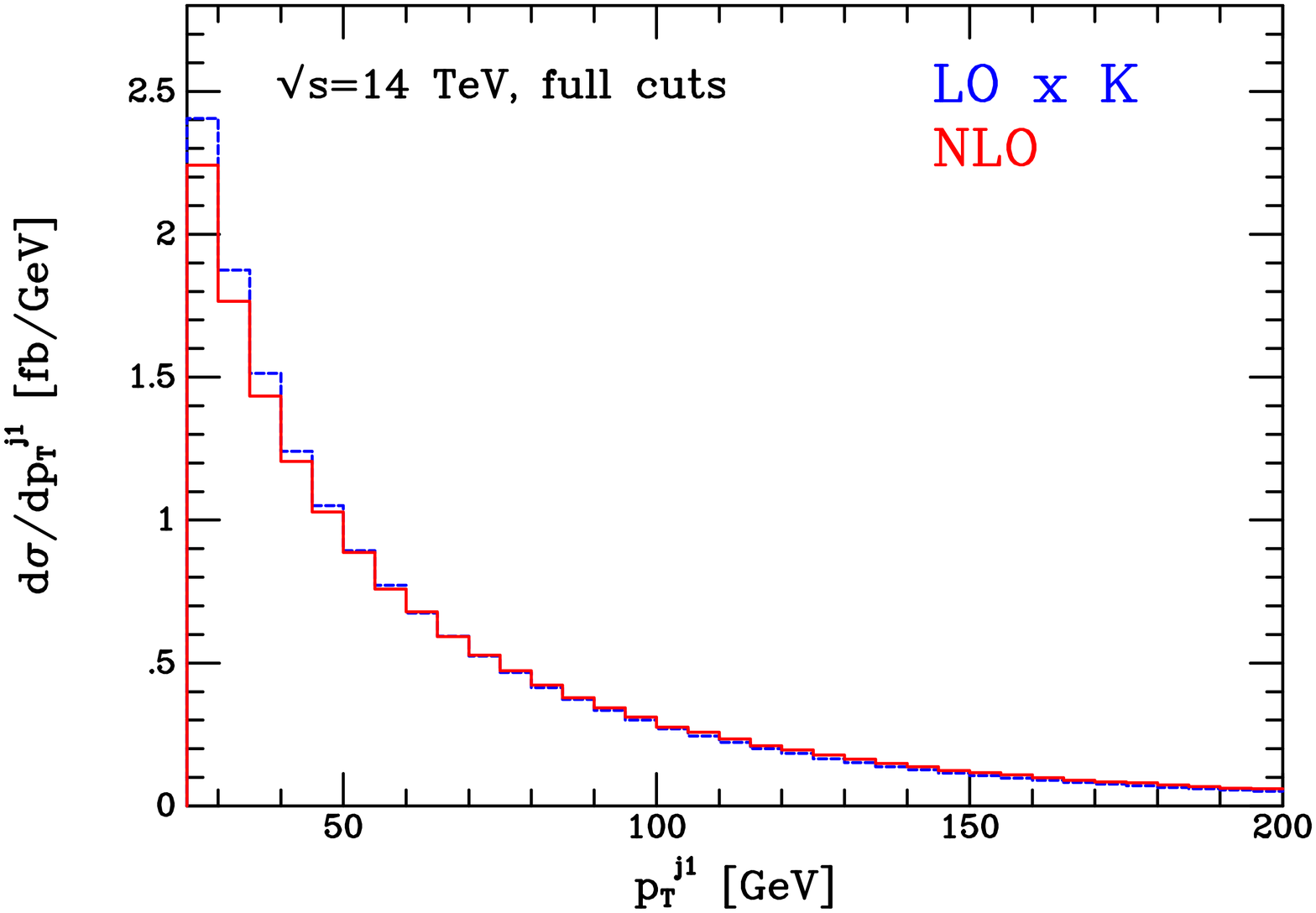}
\end{minipage}
\begin{minipage}{0.45\textwidth}
\includegraphics[height=\textwidth,angle=-90]{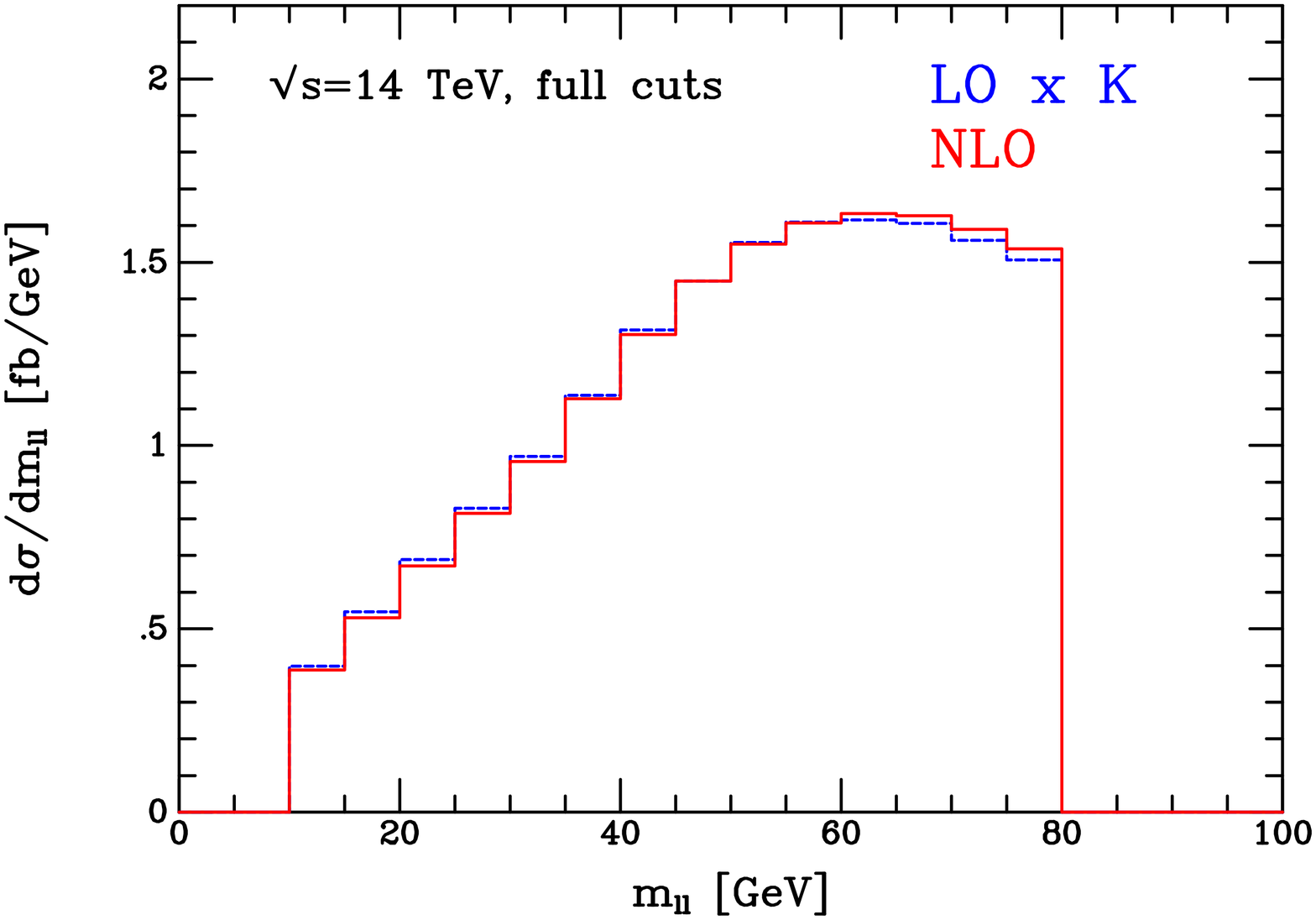}
\end{minipage}
\caption{\label{fig:14TeV} Kinematic distributions at 14 \TeV,
using the full set of cuts specified in the text.  The NLO prediction
is shown as the solid (red) histogram, while the dashed (blue) histogram
corresponds to the LO prediction rescaled by the $K$-factor from Table~\ref{14TeVresults}.}
\end{figure}

In the results that follow we shall
always consider the decay of each $W$ boson into a single lepton family, i.e.
the Born level quark-antiquark process we consider is the one shown
in Eq.~(\ref{WWjetprocess}).
The cross-sections under these cuts are given in Table~\ref{14TeVresults}.
In order to assess their effect, we also show for comparison the cross
sections obtained using only the jet cuts, i.e. the top two lines
of the cuts in Table~\ref{14TeVcuts}.  In addition, we consider the
imposition of an additional {constraint} on the transverse momentum of the
putative Higgs boson,
\begin{equation}
p_{\perp}^H\,\equiv\,\sum\,p_{T\,\text{miss}}+p_{T,\ell\ell} < 125~\rm{GeV} \;.
\label{ptHcut}
\end{equation}
Such a cut is useful when testing the spin-2 hypothesis for the Higgs
boson~\cite{Aad:2015rwa}.  The table also shows the $K$-factor, defined by $K=\sigma^{\rm{NLO}}/\sigma^{\rm{LO}}$,
which we find is rather insensitive to which set of cuts is applied.
\begin{center}
\begin{table}
\begin{tabular}{l|l|l|l}
cuts ~~~& $\sigma^{\rm{LO}}$ [fb] ~~& $\sigma^{\rm{NLO}}$ [fb]~~&~~$K$~~ \\ \hline
basic   & 462.0(2)                & 568.4(2)& 1.23 \\
full    & 67.12(4)                & 83.91(5)& 1.25 \\
spin-2  & 58.21(4)                & 71.32(5)& 1.23 \\
\end{tabular}
\caption{Cross-sections at 14 TeV, for the cuts specified in
Table~\ref{14TeVcuts} (basic, full) and also in Eq.~(\ref{ptHcut}) (spin-2).
Monte Carlo uncertainties are indicated in
parentheses and are smaller than the per mille level. \label{14TeVresults}}
\end{table}
\end{center}
Going beyond the pure cross-section calculation, it is interesting to examine
the effect of NLO corrections on a a few key differential distributions.
We shall consider a number that have already entered in the discussion
of the cuts -- $m_T^{\ell\ell}$, $\Delta \Phi_{\ell\ell}$ and $m_{\ell\ell}$ --
as well as the transverse momentum of the lead jet, $p_{\perp}^{j_1}$.
These quantities are shown in Figure~\ref{fig:14TeV} where, for comparison,
the LO prediction has been rescaled by the $K$-factor from Table~\ref{14TeVresults}.
This indicates that there is very little difference
between the shapes of the distributions at each order, with the exception
of the transverse momentum of the leading jet.  In contrast this does receive
significant corrections, which is expected since additional radiation beyond
a single jet is only present at NLO.

\begin{figure}
\begin{minipage}{0.45\textwidth}
\includegraphics[height=\textwidth,angle=-90]{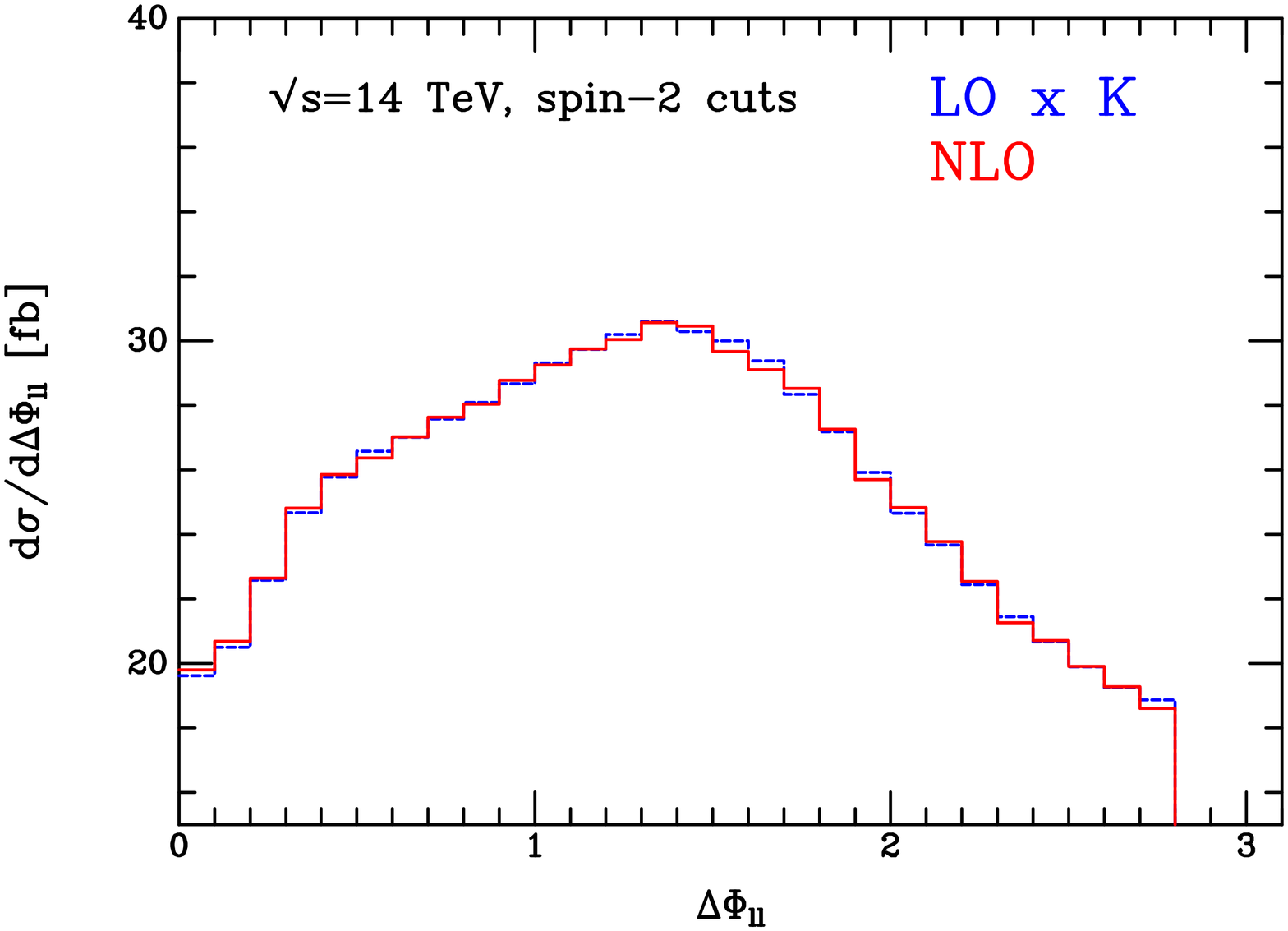}
\end{minipage}
\begin{minipage}{0.45\textwidth}
\includegraphics[height=\textwidth,angle=-90]{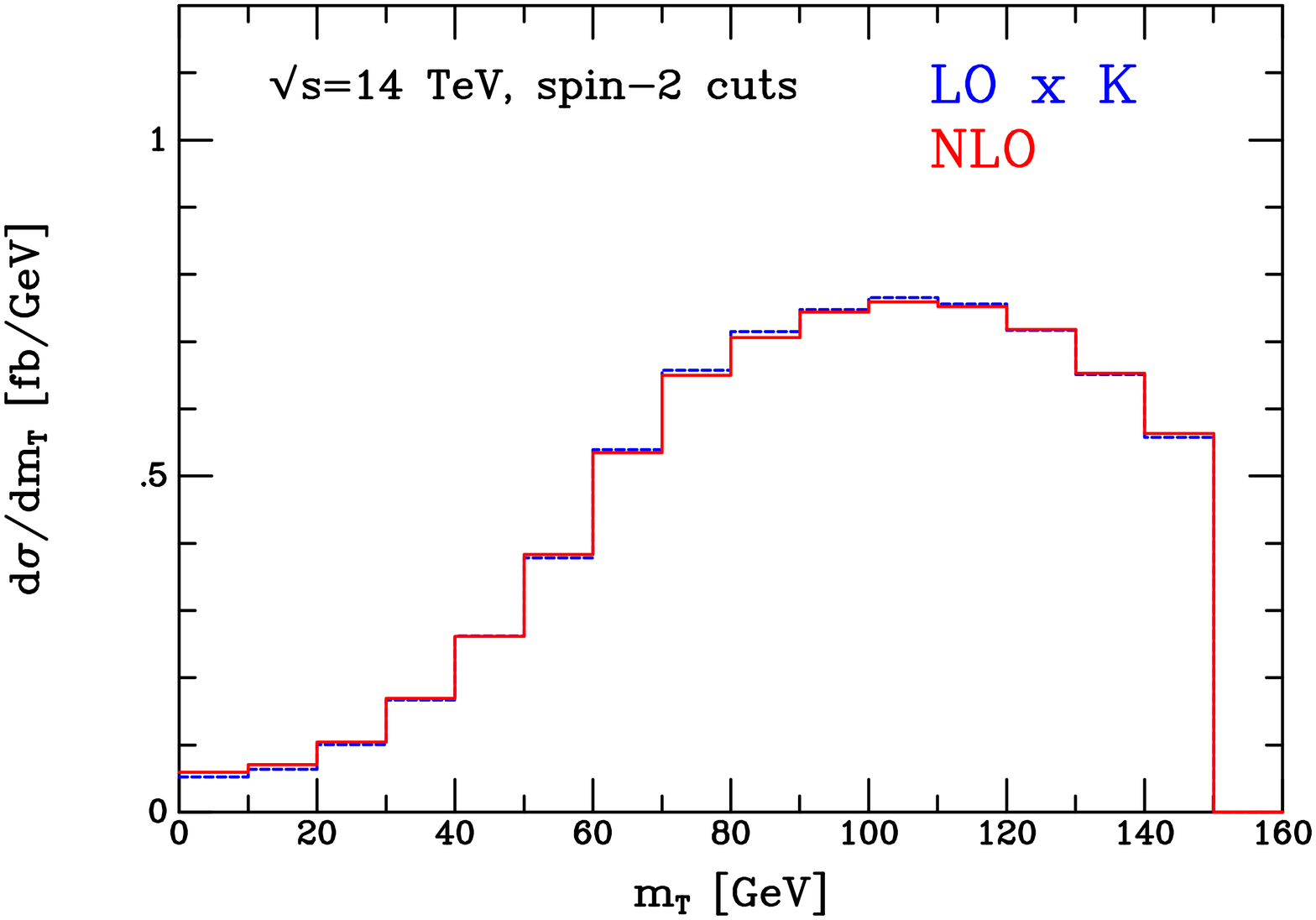}
\end{minipage}
\begin{minipage}{0.45\textwidth}
\includegraphics[height=\textwidth,angle=-90]{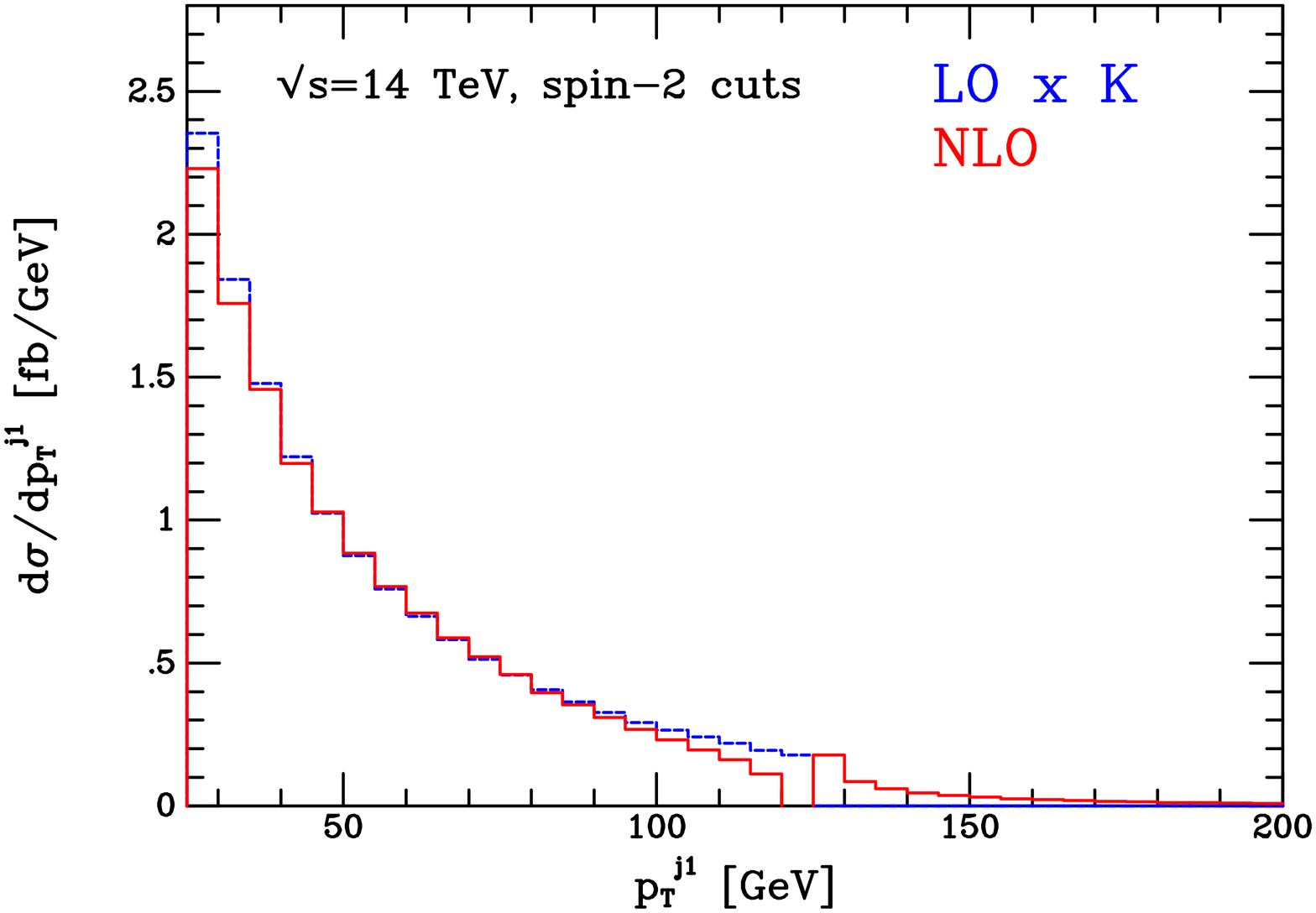}
\end{minipage}
\begin{minipage}{0.45\textwidth}
\includegraphics[height=\textwidth,angle=-90]{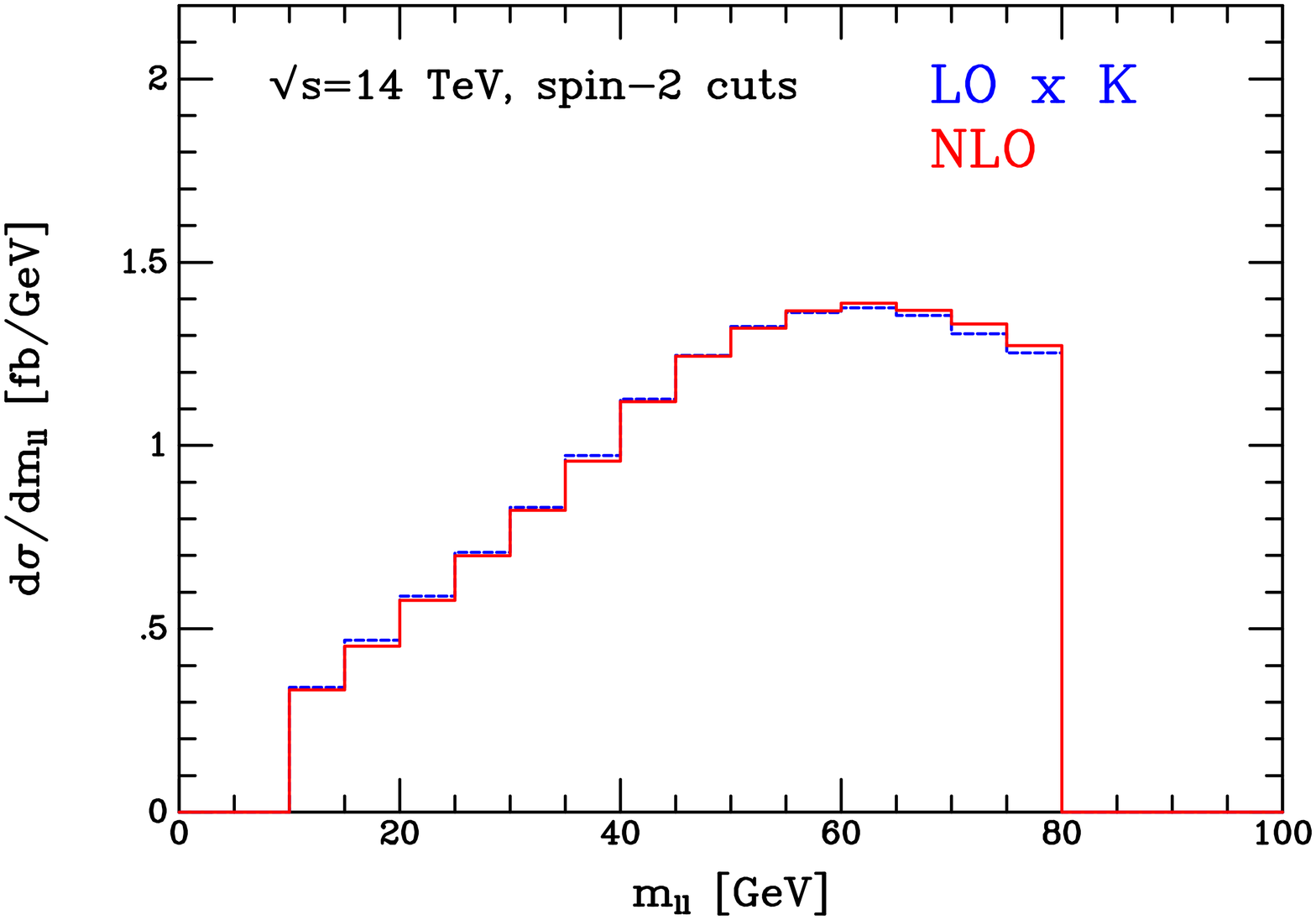}
\end{minipage}
\caption{\label{fig:14TeVptHcut} Kinematic distributions at 14 \TeV,
using the spin-2 set of cuts specified in the text. The labelling
is as in Figure~\ref{fig:14TeV}.}
\end{figure}

The corresponding distributions after the application of the spin-2
cuts, i.e. the addition of the transverse momentum cut in Eq.~(\ref{ptHcut}),
are shown in Figure~\ref{fig:14TeVptHcut}.  The additional cut has little
effect on the distributions, except for $p_{\perp}^{j_1}$.  This exhibits a
discontinuity at $125$~GeV, reflecting the fact that the NLO prediction is
not reliable in this region due to the kinematic limitation present
at LO ($p_{\perp}^{j_1} = p_{\perp}^H$).

\begin{center}
\begin{table}
\begin{tabular}{l|l}
variable& cut \\ \hline
$p_{\perp,j}$&$>$ 30 \GeV\\
$|\eta_j|$&$<4.5$ \\
\hline
$|\eta_\ell|$&$<$ 2.5 \\
$p_{\perp,\ell_1}$&$>$ 50 \GeV \\
$p_{\perp,\ell_2}$&$>$ 10\,\GeV \\
$p_\perp^\text{miss}$&$>$ 20\,\GeV \\
$m_T^{\ell\ell}$&$>$ 80 \GeV \\
\end{tabular}
\caption{\label{100TeVcuts} 
Cuts applied in the $100$~TeV analysis, corresponding to
the ``full'' set of cuts.  The jet cuts, corresponding to the
first two lines in the table, are the only ones applied for
the ``basic'' cross-section.
}
\end{table}
\end{center}

For our $100$~TeV analysis, we take as inspiration the CMS search for
additional heavy resonances presented in Ref.~\cite{Khachatryan:2015cwa}.
Such resonances can appear in models with extended Higgs sectors, where the simplest realization contains an
additional scalar which is a singlet under all SM gauge groups \cite{Schabinger:2005ei,Patt:2006fw} (see
also \cite{Pruna:2013bma, Robens:2015gla} and references therein).  This can be interpreted as a limiting
case for more generic BSM scenarios, e.g.~models with either additional gauge sectors~\cite{Basso:2010jm} or
matter content~\cite{Strassler:2006im,Strassler:2006ri}. In such models, the decay modes for the
additional heavy scalar are dominated by $WW$ decays with branching ratios of order $80\%$ or
higher~\cite{Robens:2015gla}; therefore the process considered here constitutes a dominant SM background.
Furthermore, as the parameter space of such models is severely constrained by electroweak precision
measurements \cite{Lopez-Val:2014jva}, direct production cross-sections at the LHC are typically
of order a few tens of femtobarns, even for additional scalar masses below 1 \TeV. Therefore, such
models may be hard to constrain at the LHC and would remain to be investigated at a future $100$~TeV
proton-proton collider. 

\begin{figure}
\begin{minipage}{0.45\textwidth}
\includegraphics[height=\textwidth,angle=-90]{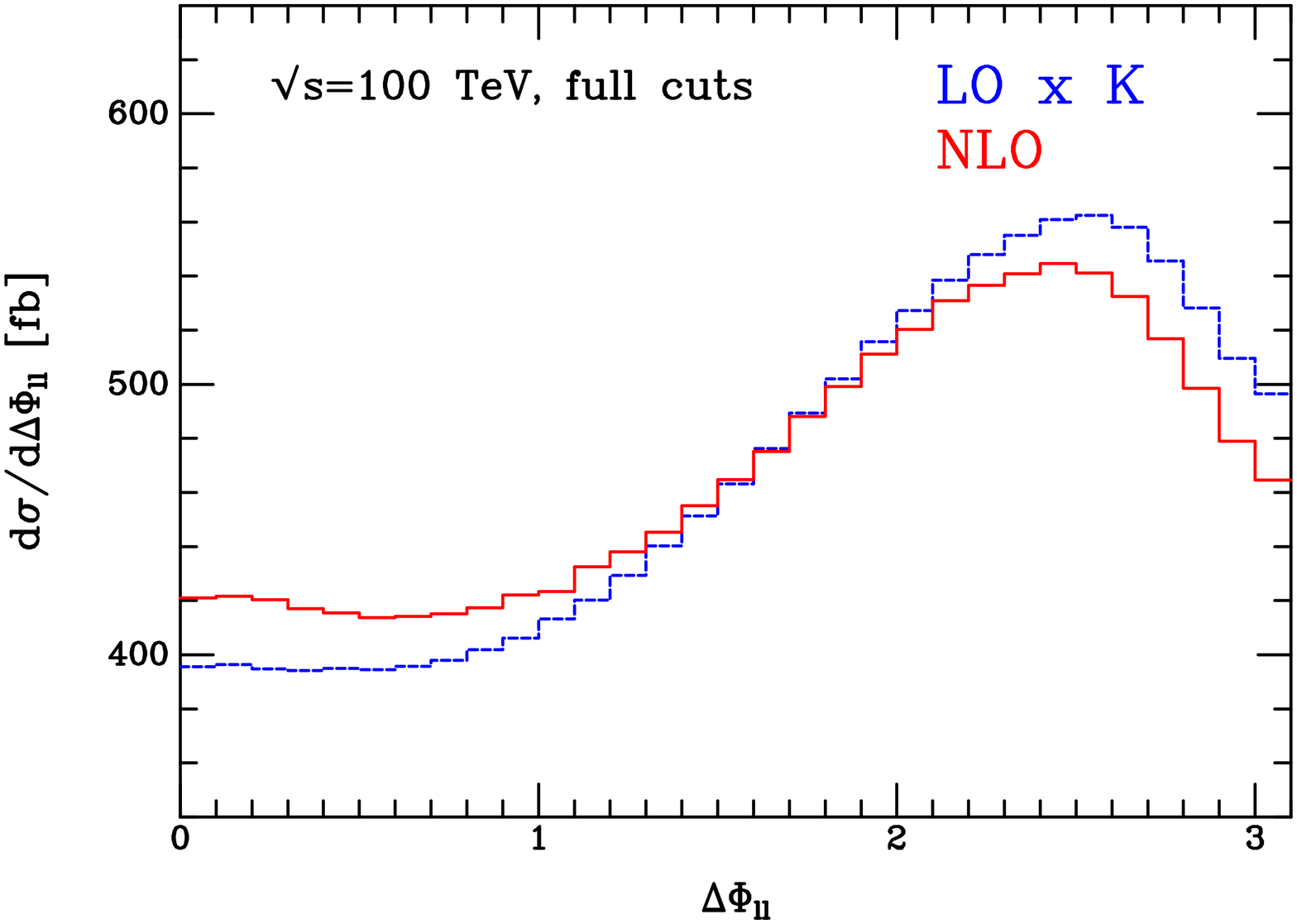}
\end{minipage}
\begin{minipage}{0.45\textwidth}
\includegraphics[height=\textwidth,angle=-90]{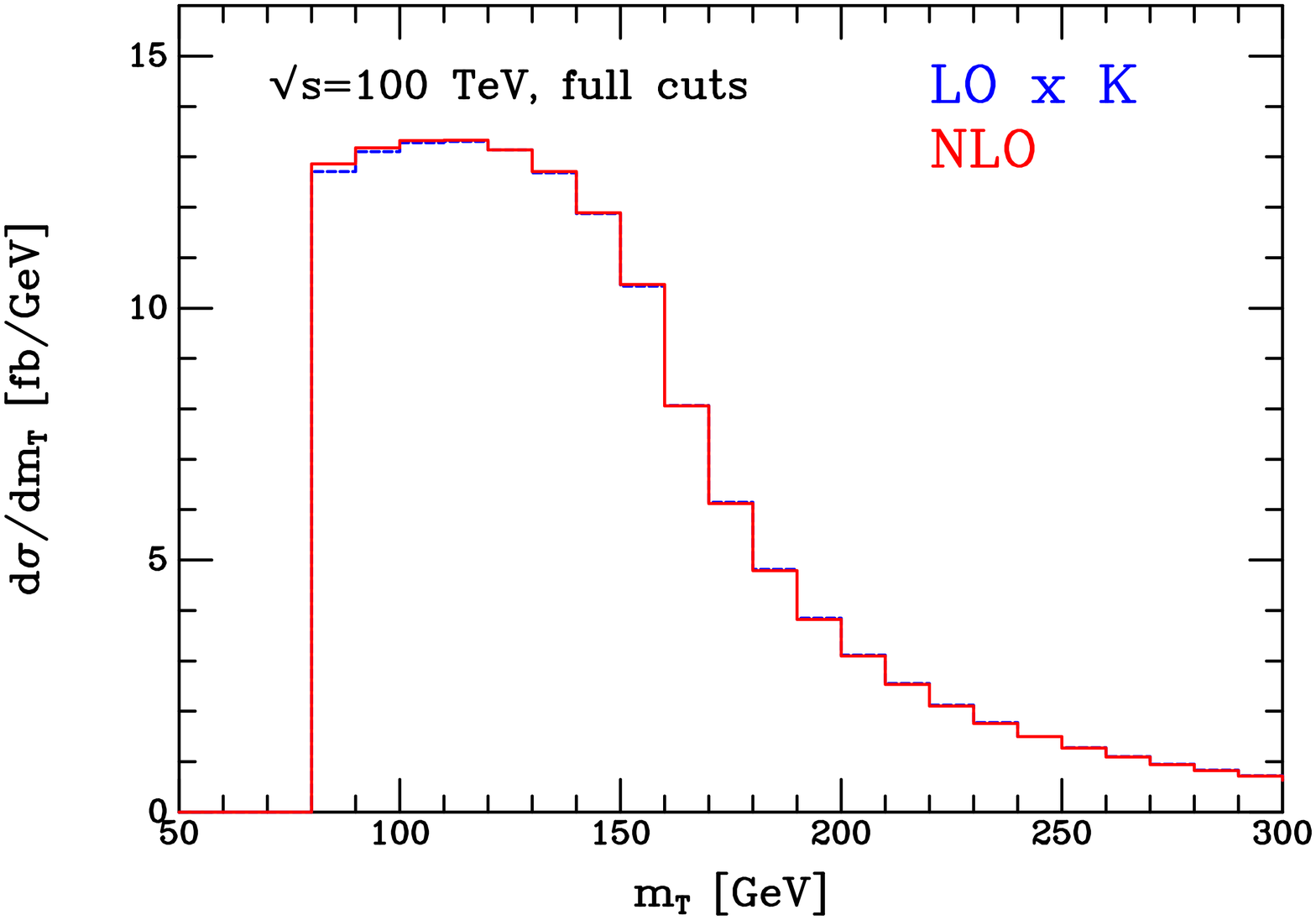}
\end{minipage}
\begin{minipage}{0.45\textwidth}
\includegraphics[height=\textwidth,angle=-90]{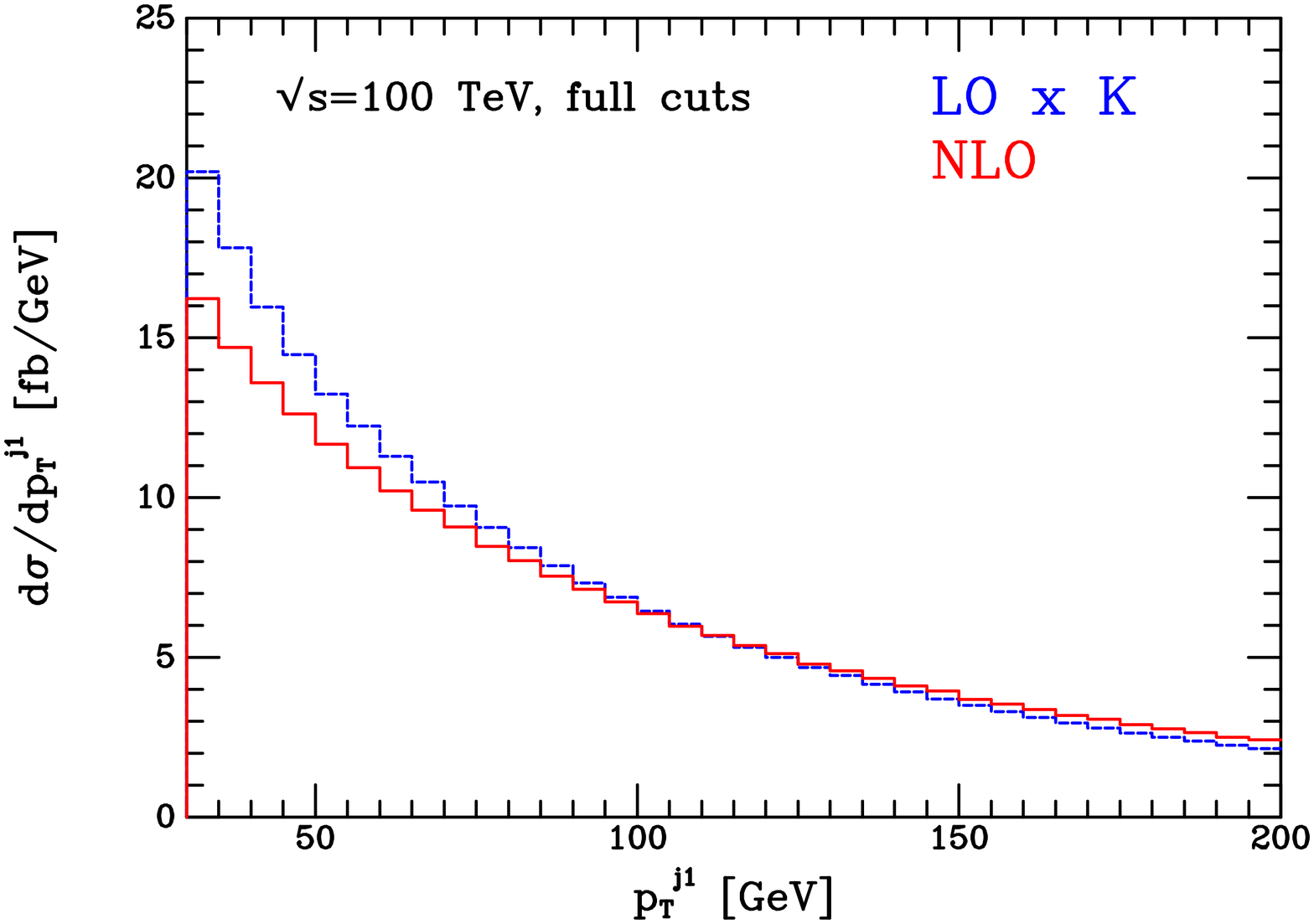}
\end{minipage}
\begin{minipage}{0.45\textwidth}
\includegraphics[height=\textwidth,angle=-90]{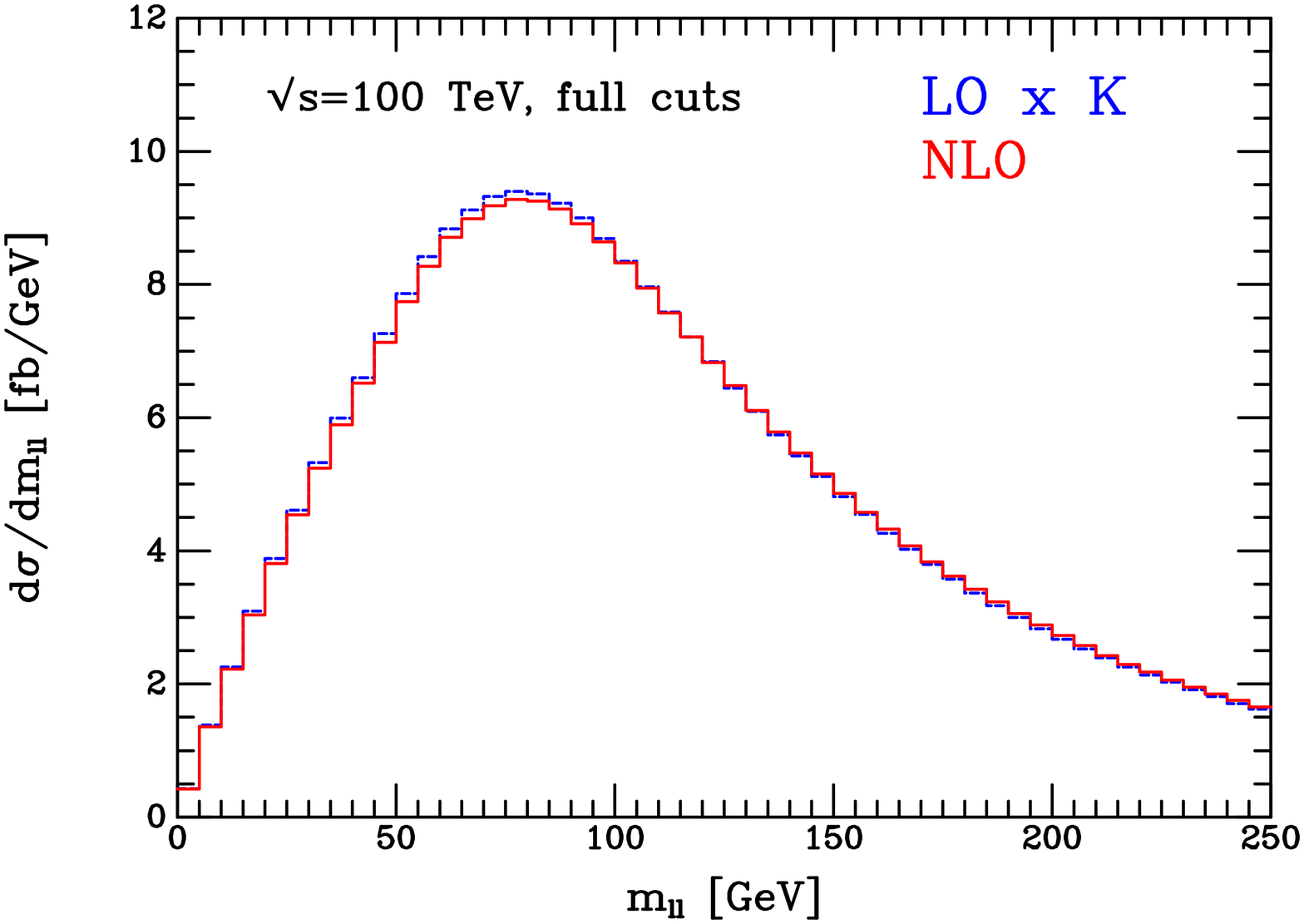}
\end{minipage}
\caption{\label{fig:100TeV} Kinematic distributions at 100 \TeV,
using the full set of cuts specified in the text. The labelling
is as in Figure~\ref{fig:14TeV}.}
\end{figure}

The cuts for our $100$~TeV analysis are shown in Table~\ref{100TeVcuts}, where we have adopted the $p_\perp$ and
$m_T$ cut values used in searches for heavy resonances with masses greater than $200\,\GeV$.\footnote{Note
that in Ref.~\cite{Khachatryan:2015cwa}, the $m_T^{\ell\ell}$ cut at $80\,\GeV$ in the case of heavy resonances is
erroneously reported as a cut on $E_{\perp,\text{miss}}$. We thank X. Janssen for clarifying this point. }
For the sake of simplicity we do not adopt the separation into strict jet bins performed in Ref.~\cite{Khachatryan:2015cwa},
nor do we discard two-jet events that fail the vector boson fusion selection cuts.\footnote{
The vector boson fusion selection cuts correspond to $m_{jj}>500\,\GeV$ and $\Delta\eta_{jj}>3.5$~\cite{Khachatryan:2015cwa}.
In our case such cuts can only apply to real radiation contributions that are present at NLO. We find that these amount to
about $90\,\fb$, which can be used as an order of magnitude estimate of contamination for the pure
VBF-type signal from quark-induced $WW\,+\,2\,\text{jets}$ at leading order.}   
The cross-sections at $100$~TeV are shown in Table~\ref{100TeVresults}. We again find that $K$-factors do
not depend strongly on the choice of cuts. In Figure~\ref{fig:100TeV} we again display
differential distributions in our usual kinematic quantities,  $m_T^{\ell\ell}$, $\Delta \Phi_{\ell\ell}$,
$m_{\ell\ell}$, and $p_{\perp}^{j_1}$, showing both NLO as well as rescaled LO results. We observe that
a correct description of the $\Delta \Phi_{\ell\ell}$ and $p_{\perp}^{j_1}$ distributions requires
the inclusion of the full NLO corrections. On the other hand, the distributions for $m_T^{\ell\ell}$ and
$m_{\ell\ell}$ are both well described using a rescaled LO calculation.
\begin{center}
\begin{table}
\begin{tabular}{l|l|l|l}
cuts ~~~& $\sigma^{\rm{LO}}$ [pb] ~~& $\sigma^{\rm{NLO}}$ [pb]~~&~~$K$~~ \\ \hline
basic   & 6.815(1)                  & 7.939(5)& 1.16 \\
full    & 1.237(1)                  & 1.471(1)& 1.19 \\
\end{tabular}
\caption{ cross-sections at 100 TeV, for the cuts specified in
Table~\ref{100TeVcuts}. Monte Carlo uncertainties are indicated in
parentheses and are smaller than the per mille level. \label{100TeVresults}}
\end{table}
\end{center}

\section{Conclusions}
In this paper we have presented a calculation of $WW$ production in association with a jet at
next-to-leading order in QCD using generalized unitarity methods.  These methods allow the one-loop
amplitude to be determined in a relatively compact analytic form.  As examples,  we have shown several
expressions for the coefficients of box, triangle, and bubble scalar integrals that appear in the
amplitudes. Our calculation has been implemented in the parton level Monte Carlo
generator MCFM, which contains analytic expressions for the complete virtual amplitude.
The MCFM code is significantly faster than previous publicly
available implementations of the one-loop amplitudes.
To demonstrate the effect of including NLO corrections,
we have studied several cases of phenomenological interest where the $WW$+jet process serves as an important
background.  These included studies of the CP properties and spin of the recently-discovered Higgs boson at the LHC, as
well as searches for additional scalar resonances at a $100$~TeV proton-proton
collider. We have provided cross-section predictions for both scenarios, and examined a number of kinematic
distributions relevant for the experimental analyses. For several of these, and especially the $p_\perp$
distribution of the first jet, the
fixed order NLO description is significantly different from the prediction obtained at LO.

Apart from the phenomenological studies, the full analytic expression for the virtual amplitude is an
important ingredient in the determination of the $WW$ production cross-section at next-to-next-to-leading
order. These contributions will be made available in the release of the next version of the MCFM
code.

\section*{Acknowledgements}
TR thanks Simon Badger, Ruth Britto, Fabrizio Caola, Pierpaolo Mastrolia, and Ciaran Williams for
extremely useful discussions during this work, as well as the Fermilab theory group for their
hospitality. DJM is supported by the UK Science and Technology Facilities Council (STFC) under grant ST/L000446/1. DJM and TR also want to thank A.A.H. Graham for contributions during the early stages of this work. This research is supported by the US DOE under contract DE-AC02-07CH11359.

\bibliography{paper}
\end{document}